\documentclass[12pt,twoside,a4paper]{article}
\usepackage{amsmath,amssymb,latexsym,theorem,natbib,epsfig,color,subfigure,footmisc,bbm}
\usepackage{multirow,graphicx,array,rotating,url}  
\usepackage{epstopdf,afterpage,wasysym}
\usepackage{verbatim}
\usepackage[table]{xcolor}
\usepackage{wrapfig}
\usepackage{placeins}
\usepackage{natbib}
\usepackage{hyperref}
\usepackage{booktabs}
\usepackage[normalem]{ulem}
\numberwithin{equation}{section}

\def\rank{\mathop{\hbox{\rm rank}}}

\title{\Large A Composite-Loss Graph Neural Network for the Multivariate Post-Processing of Ensemble Weather Forecasts}

\author{\large M\'aria Lakatos \\
\normalsize \url{lakatos.maria@inf.unideb.hu}\vspace{0.5cm} \\
\normalsize Faculty of Informatics, University of Debrecen, Hungary
}

\date{}

\begin{document}

\maketitle

\begin{abstract}
Ensemble forecasting systems have advanced meteorology by providing probabilistic estimates of future states, supporting applications from renewable energy production to transportation safety. Accurate and reliable forecasts are critical for operational optimization, system stability, and informed decision-making. Nonetheless, systematic biases often persist, making statistical post-processing essential. Traditional parametric post-processing techniques and machine learning-based methods can produce calibrated predictive distributions at specific locations and lead times, yet often struggle to capture dependencies across forecast dimensions. To address this, multivariate post-processing methods—such as ensemble copula coupling and the Schaake shuffle—are widely applied in a second step to restore realistic inter-variable or spatio-temporal dependencies.

The aim of this study is the multivariate post-processing of ensemble forecasts using a graph neural network (dualGNN) trained with a composite loss function that combines the energy score (ES) and the variogram score (VS). The method is evaluated on two datasets: WRF-based solar irradiance forecasts over northern Chile and ECMWF visibility forecasts for Central Europe. The dualGNN consistently outperforms all empirical copula-based post-processed forecasts and shows significant improvements compared to graph neural networks trained solely on either the continuous ranked probability score (CRPS) or the ES, according to the evaluated multivariate verification metrics. Furthermore, for the WRF forecasts, the rank-order structure of the dualGNN forecasts captures valuable dependency information, enabling a more effective restoration of spatial relationships than either the raw numerical weather prediction ensemble or historical observational rank structures. Notably, incorporating VS into the loss function improved the univariate performance for both target variables compared to training on ES alone. Moreover, for the visibility forecasts, the ES–VS combination even outperformed the strongest calibrated reference in terms of univariate performance.

\bigskip

\end{abstract}

\section{Introduction}
\label{sec1}

Ensemble forecasts, generated by combining multiple model runs with varied initial conditions or configurations, provide a probabilistic framework that captures forecast uncertainty--an essential component for informed decision-making across a wide range of applications. Despite their advantages, these ensemble outputs often require calibration to correct for systematic biases \citep[see e.g.][]{hamill1997verification}. Statistical post-processing aims to produce calibrated forecasts that ensure consistency between observations and predictions, thereby enhancing probabilistic accuracy and overall reliability. This process is particularly important in regions where accurate environmental forecasts are critical for decision-making and the management of infrastructure. 

A wide range of statistical post-processing methods has been developed to improve the calibration of ensemble forecasts; for a comprehensive overview, we refer the reader to \citet{vannitsem2021pp}. Among the most prominent are ensemble model output statistics \citep[EMOS;][]{g2005} and Bayesian model averaging \citep[BMA;][]{r200}, both of which link ensemble forecasts to parametric predictive distributions and have been successfully applied to various meteorological variables. More recently, machine learning approaches have received increasing attention in this domain. Distributional regression networks \citep[DRNs;][]{rl18}, for instance, directly map ensemble inputs to full predictive distributions and have demonstrated improved performance compared with traditional statistical methods across multiple applications. More advanced architectures--such as transformers, convolutional neural networks, generative adversarial networks, and distributional regression UNets--have subsequently been applied to post-process forecasts across a wide range of atmospheric variables \citep[see e.g.][]{bouallegue2024improving, li2022convolutional, dai2021spatially, pic2025distributional}.

A widely recognized limitation of most post-processing methods is that they are typically applied independently for each forecast horizon, location, and variable, potentially neglecting important dependencies across these dimensions. Multivariate post-processing aims to recover the dependence structure between different forecast dimensions, often lost in univariate calibration. This can be achieved using parametric approaches, such as Gaussian copulas \citep{moller2013multivariate}, or non-parametric copula-based methods, which reconstruct coherent joint predictive distributions without requiring a specific parametric form. These approaches are often termed two-step methods: first, samples are generated from the calibrated predictive distributions, and then they are rearranged according to a dependence template. Examples include ensemble copula coupling \citep{schefzik2013uncertainty} and the Schaake shuffle \citep{clark2004schaake}. Machine learning-based approaches are increasingly applied in multivariate post-processing as well. In particular, generative models have shown promise in producing spatially coherent forecast scenarios, offering a flexible alternative to traditional copula-based techniques by directly learning complex dependency structures from data \citep{chen2024}.

Graph neural networks (GNNs) are an emerging approach in machine learning, yet their application to the post-processing of ensemble forecasts remains limited. A major strength of GNNs is their ability to model spatial dependencies, which is crucial for accurate meteorological forecasting. While \cite{feik2024graph} employed a GNN to generate calibrated forecast distributions for 2-m temperature predictions, more recently \cite{bulte2025graph} applied GNNs to improve extreme rainfall forecasts.

In this study, we employ a graph neural network (GNN) to post-process ensemble forecasts, generating improved ensemble predictions directly at the output layer rather than estimating the parameters of the forecast distributions. Our focus is on solar irradiance and visibility, but the method is fully general and non-parametric, making it applicable to any meteorological variable and flexible with respect to the size of the post-processed ensemble. We investigate several variants of the GNN trained with different loss functions: one minimizing the continuous ranked probability score (CRPS), and another trained on the energy score (ES) following the approach of the conditional generative model by \citet{chen2024}. Our primary objective is to evaluate whether a GNN trained with a composite loss combining the ES and the variogram score (VS) can deliver improved predictive performance by simultaneously preserving overall distributional accuracy and better capturing spatial dependencies.

The paper is organized as follows. Section 2 provides a brief description of the solar irradiance and visibility datasets studied. In Section 3, we review the univariate and multivariate post-processing methods applied, as well as the verification metrics used to evaluate forecast performance. Section 4 presents the details of the model implementations. The results of our two case studies are reported in Section 5, followed by a brief discussion and concluding remarks in Section 6.

\section{Data}
\label{sec2}

\begin{figure}
\centering
  \includegraphics[width=.8\linewidth]{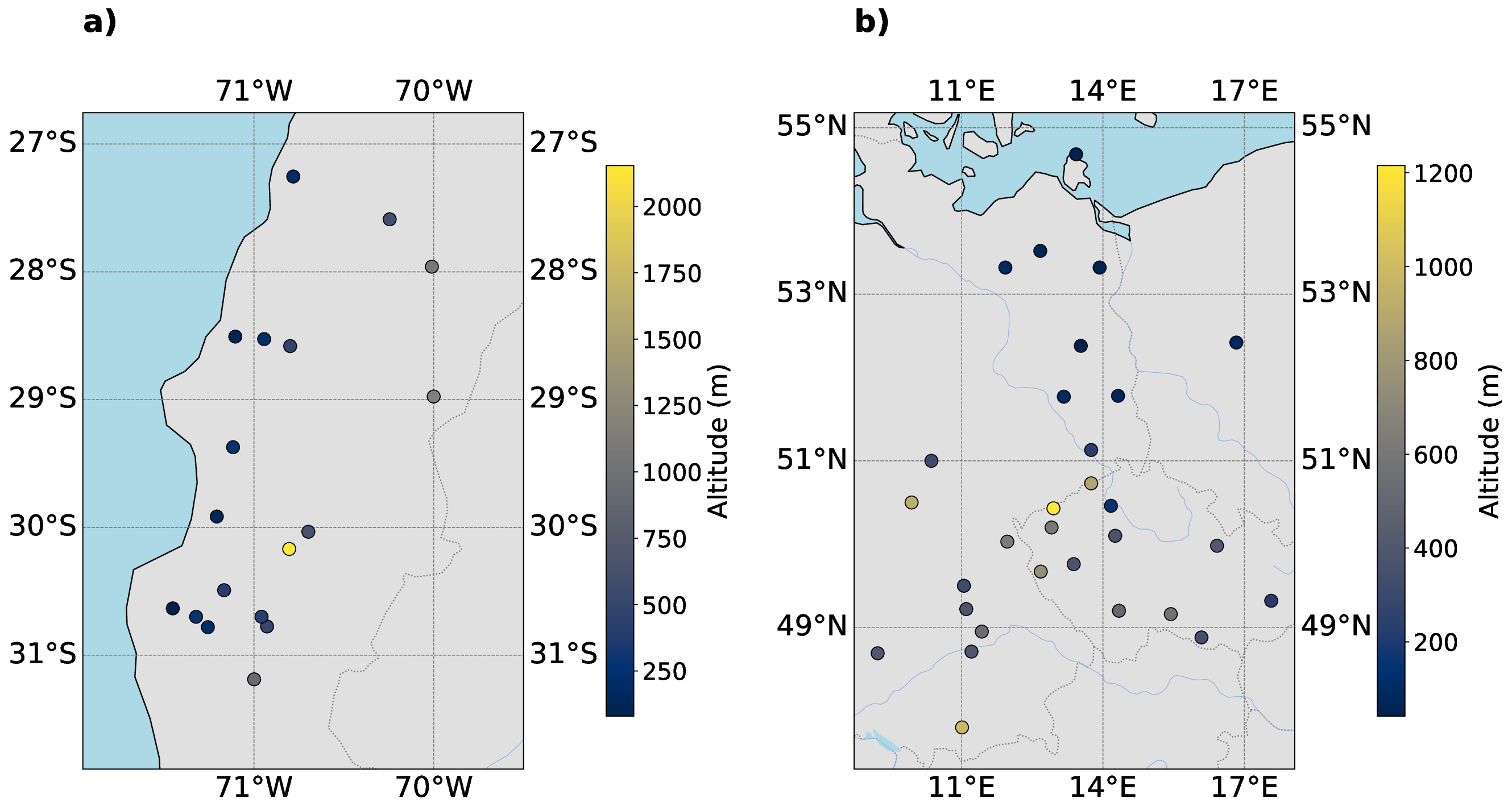}
  \caption{Location of solar irradiance observation stations in northern Chile (a) and visibility observation stations across Central Europe (b).}
  \label{fig:chilemap}
\end{figure}

The solar irradiance forecasts utilized here --also invsetigated by \citet{bmcdsznl25}--were generated with the Weather Research and Forecasting (WRF) model (version 4.4.2; \citealp{skamarock2019}) for the year 2021. An eight-member ensemble was employed, with a 3~km horizontal resolution and 36 vertical levels, capturing the key atmospheric layers. The model domain covers a wide range of Chilean landscapes, from the Pacific coast to the Andes. Observations, expressed in \(W/\mathrm{m}^{2}\), were obtained from the National Weather Service’s monitoring network (\url{https://climatologia.meteochile.gob.cl/}). After excluding stations with more than 10\% missing data, 18 stations in the Atacama and Coquimbo regions (Figure \ref{fig:chilemap}a) remained for model training and evaluation. All forecasts were initialized at 0000 UTC, with a 1 h temporal resolution and a forecast horizon of 48 h.

The second case study examines visibility forecasts from the European Centre for Medium-Range Weather Forecasts (ECMWF) at 30 SYNOP stations across Germany, the Czech Republic, and Poland (Figure \ref{fig:chilemap}b) for 2020–2021. Forecasts were initialized daily at 0000 UTC, with 20 lead times ranging from 6 h to 120 h at 6 h intervals. Each ensemble includes the operational control run with unperturbed initial conditions, together with 50 additional members produced by the ECMWF Integrated Forecasting System (IFS), which are statistically indistinguishable, hence exchangeable.  Although ECMWF visibility forecasts are issued in 1 m increments and can therefore be treated as continuous, SYNOP stations typically report observations in discrete categories specified by WMO recommendations—namely, "100 to 5000 m in steps of 100 m, 6 to 30 km in steps of 1 km, and 35 to 70 km in steps of 5 km" \citep[Section 9.1.2]{wmo2018guide}. Accordingly, in Sections \ref{sec:uv_results_vis} and \ref{sec:mv_results_vis}, the post-processed visibility forecasts are evaluated with respect to these 84 categories.

\section{Methods for post-processing and forecast evaluation}
\label{sec3}

This section details the univariate post-processing models applied to the solar irradiance and visibility ensemble forecasts. Section \ref{sec3.1} describes the EMOS model used for solar irradiance forecasts, along with a multilayer perceptron (MLP) originally proposed by \cite{bmcdsznl25}. Section \ref{sec3.2} describes the classifier employed for visibility forecast post-processing. While \cite{lakatos2024enhancing} also applied an MLP to the raw forecasts, the proportional odds logistic regression (POLR) model generally achieved  better performance and is therefore adopted here as the reference model.

\subsection{Univariate post-processing of solar irradiance}	
\label{sec3.1}

When modeling solar irradiance, it is necessary to account for its non-negativity and the frequent occurrence of zero observations; in such cases, probability distributions that assign a positive mass to zero are appropriate. Following the approaches of \cite{bmcdsznl25} and \cite{schulz2021}, we employ the censored normal EMOS model, which enables a proper probabilistic representation of both positive and zero irradiance values.

For parameter estimation, we adopt the semi-local EMOS approach of \cite{lerch2017similarity}, which clusters stations with similar climatological and/or forecast error characteristics. Within each cluster, all stations share a common set of EMOS parameters, jointly estimated from the data of the stations in that cluster. This methodology facilitates more robust parameter estimation while preserving local representativeness, in contrast to the global approach \citep{thorarinsdottir2010probabilistic} that fits a single parameter set to data from all stations, potentially oversmoothing spatial variability.
\bigskip

Following the notations of \cite{bmcdsznl25}, let $G(x \mid \mu, \sigma) := \Phi\big((x-\mu)/\sigma\big)$, $x \in \mathbb{R}$, denote the cumulative distribution function (CDF) of a Gaussian distribution with mean $\mu$ and standard deviation $\sigma>0$, where $\Phi$ is the standard normal CDF. The CDF of a Gaussian distribution with location $\mu$ and scale $\sigma$ left-censored at zero is then
\[
G^c_0(x \mid \mu, \sigma) :=
\begin{cases}
G(x \mid \mu, \sigma), & x \ge 0, \\
0, & x < 0.
\end{cases}
\]
This distribution assigns a probability mass of $G^c_0(0 \mid \mu, \sigma)$ at zero and has a mean given by
\[
\kappa = \mu \, \Phi\Big(\frac{\mu}{\sigma}\Big) + \sigma \, \phi\Big(\frac{\mu}{\sigma}\Big),
\]
where $\phi$ denotes the standard normal probability density function (PDF). Following \cite{schulz2021}, \cite{bb24} and \cite{bmcdsznl25}, the ensemble statistics are linked to the parameters of the censored normal distribution through
\[
\mu = \gamma_0 + \gamma_1 \bar{f} + \gamma_2 p_0, \qquad
\sigma = \exp\!\big(\delta_0 + \delta_1 \log S\big),
\]
where $p_0$ and $S^{2}$ denote the proportion of zero observations and the ensemble variance, respectively, defined as
\[
p_0 := \frac{1}{K} \sum_{k=1}^{K} \mathbf{1}\{f_k = 0\}, 
\qquad 
S^{2} := \frac{1}{K-1} \sum_{k=1}^{K} (f_k - \bar{f})^{2},
\]
where $\bar{f}$ is the ensemble mean and $\mathbf{1}\{\cdot\}$ denotes the indicator function.
\bigskip

An alternative to conventional statistical post-processing methods is the application of machine learning models that directly produce calibrated ensemble forecasts. In this study, the multilayer perceptron (MLP) approach of \cite{bmcdsznl25} is adopted, in which the number of output neurons is set equal to the target ensemble size. The network is trained by minimizing the sample CRPS (\ref{eq:sample_crps}), under the constraint that predicted solar irradiance values are non-negative. This distribution-free, data-driven framework allows flexible specification of the ensemble size, provided sufficient training data are available. To enable a fair comparison with the raw WRF forecasts, eight-member ensembles are generated for each prediction case. Previous experiments on the Chilean dataset have demonstrated that this approach outperforms the distributional regression network (DRN), which instead predicts the location $\mu$ and scale $\sigma$ parameters of a left-censored normal distribution. In what follows, this method is referred to as MLP.

\subsection{Univariate post-processing of visibility}
\label{sec3.2}

As mentioned in the Introduction, visibility observations reported by the WMO form a discrete set of categories \citep{wmo2018guide}; as a result, the post-processing of visibility forecasts can be viewed as a classification problem. Following \cite{lakatos2024enhancing}, the calibration of visibility ensemble forecasts is performed using proportional odds logistic regression \citep[POLR;][]{mccullagh1980regression}, a widely used parametric method for modeling ordinal response variables, which makes it particularly well suited for the visibility observations considered here, and thus serves as a powerful benchmark. The model specifies the conditional cumulative distribution function of the observed visibility $Y$ given an $M$-dimensional feature vector $\mathbf{x}$ as
\[
P\!\big(Y \leq y_i \mid \mathbf{x}\big) = \frac{\exp\!\big(L_i(\mathbf{x})\big)}{1 + \exp\!\big(L_i(\mathbf{x})\big)}, 
\qquad 
L_i(\mathbf{x}) := \alpha_i + \mathbf{x}^\top \boldsymbol{\beta}, 
\qquad 
i = 1, 2, \ldots, 84,
\]
where $\alpha_i \in \mathbb{R}$ and $\boldsymbol{\beta} \in \mathbb{R}^{M}$ are model parameters satisfying $\alpha_1 < \alpha_2 < \cdots < \alpha_{84}$. Consequently, fitting a POLR model to the visibility data requires estimating $84 + M$ parameters. In the present work, we apply the local POLR model, as also used by \cite{lakatos2024enhancing}, for the calibration of visibility forecasts. In their study, this model provided the best predictive performance in terms of univariate performance across various spatial selection procedures, including semi-local and regional estimates (briefly discussed in Section \ref{sec3.1}), where local indicates that a separate model is fitted for each station based solely on its own historical data.

\subsection{Multivariate post-processing methods}
To restore any potentially lost spatial dependencies, this study considers two-step approaches for multivariate post-processing as benchmark models. These benchmarks generate multivariate samples by combining forecasts that have been individually calibrated in a univariate manner using an empirical copula. Several variants exist depending on the strategy employed to define the dependence template for restoring the dependencies. Extending the notation introduced in the previous section, let
\[
\boldsymbol{f}^{(d)} = \big(f_1^{(d)}, f_2^{(d)}, \ldots, f_K^{(d)}\big)
\] 
denote a $K$-member ensemble forecast for station $d$ ($d = 1, 2, \ldots, D$) at a given time point and lead time.
\bigskip

The ensemble copula coupling \citep[ECC;][]{schefzik2013uncertainty} leverages the rank order structure of the raw ensemble forecasts, to retain the intricate dependency patterns present within the ensemble. Adopting the notation of \citet{lakatos2023comparison}, the steps of this iterative procedure can be summarized as follows:

\begin{enumerate}
    \item For each dimension \(d\), generate a sample  $\boldsymbol{\widehat{f}}^{(d)}$ of size \(K\) from the calibrated marginal predictive distribution, arranged in ascending order.
    \item Define permutations 
    \[
    \boldsymbol{\pi}_d = \big(\pi_d(1), \pi_d(2), \ldots, \pi_d(K)\big)
    \]
    of the set \(\{1, 2, \ldots, K\}\) corresponding to the rank order of the raw ensemble forecasts 
    \(\boldsymbol{f}^{(d)} \), where 
    \(\pi_d(k) := \rank(f_k^{(d)})\), with ties resolved randomly. The ECC-calibrated sample 
$\boldsymbol{\widetilde{f}}^{(d)}$
    for dimension \(d\) is obtained by rearranging the sample from step 1 according to \(\boldsymbol{\pi}_d\), that is,
    \[
    \widetilde{f}_k^{(d)} := \widehat{f}_{\pi_d(k)}^{(d)}, \quad k=1, 2, \ldots, K, \quad d=1, 2, \ldots, D.
    \]
\end{enumerate}

A further nonparametric approach for multivariate post-processing examined here is the Schaake shuffle \citep[SSh;][]{clark2004schaake}, which reconstructs dependence structures by reordering calibrated univariate samples to match the rank order of randomly selected historical observations of the same size. Here, we limit the sample size to that of the raw ensemble and apply the same sampling procedure for comparability with ECC; however, the method can generate ensembles of arbitrary size given a sufficiently long historical record. In this study, for both case studies, we consider the equidistant quantiles of the predictive distributions as input for both the ECC and the SSh methods.

\subsubsection{Graph Neural Networks}

Graph neural networks (GNNs) provide a flexible framework for learning from data defined on irregular relational structures and have been successfully applied across diverse domains, including modeling molecular structures and predicting molecular properties \citep{li2025kolmogorov}, as well as optimizing traffic networks and route planning \citep{JIANG2022117921}. Unlike traditional neural networks such as convolutional neural networks (CNNs), which assume inputs on regular grids or lattices (e.g. image pixels), GNNs operate directly on graphs, enabling them to capture complex interactions among nodes arranged arbitrarily and non-uniformly \citep{feik2024graph}.

A graph is formally represented as 
\(\mathcal{G} = (\mathcal{V}, \mathcal{E}),\)
where the node set 
\(\mathcal{V} = \{1, 2, \ldots, D\}\)
corresponds to entities of interest, and the edges \(\mathcal{E}\) encode relationships or interactions between them. Each node \(d \in \mathcal{V}\) is associated with an $M$-dimensional feature vector 
\(\mathbf{x}^{(d)} \in \mathbb{R}^M\), 
which may include relevant covariates, measurements, or summary statistics.
 By propagating and aggregating information along the graph edges, GNNs can capture both local and global dependencies, making them a versatile tool for tasks involving structured relational data.

\subsection{Verification metrics}
Within the censored normal EMOS framework outlined in Section \ref{sec3.1}, model parameters are obtained by minimizing the mean of a proper scoring rule over the training dataset. The most widely adopted choice is the continuous ranked probability score (CRPS; \citep[Section 9.5.1]{wilks2019statistical}, a standard verification metric extensively used in the atmospheric sciences.

For a predictive CDF \(F\) and an observation \(y \in \mathbb{R}\), the CRPS is defined as
\begin{equation*}
    \mathrm{CRPS}(F, y) = \int_{-\infty}^{\infty} \left[ F(x) - \mathbf{1}\{x \ge y\} \right]^2 dx
    = \mathbb{E}|X - y| - \frac12 \mathbb{E}|X - X'|,
\label{eq:crps_general}
\end{equation*}
where \(X\) and \(X'\) are independent random variables with distribution \(F\) and finite first moments. Smaller CRPS values indicate superior predictive performance, as the score simultaneously captures calibration (statistical consistency between forecasts and observations) and sharpness (concentration of the predictive distribution). For the censored normal distribution, the CRPS admits a closed-form expression, allowing computationally efficient optimization and making it particularly suitable as a loss function in both EMOS and machine learning frameworks.

Moreover, for an ensemble forecast $\{f_1,\dots,f_K\}$ resulting in an empirical CDF  $\hat{F}_K$ the above reduces to the sample CRPS, that is
\begin{equation}
    \mathrm{CRPS}(\hat{F}_K, y) = \frac1K \sum_{k=1}^K |f_k - y| - \frac{1}{2K^2} \sum_{k=1}^K \sum_{\ell=1}^K |f_k - f_\ell|,
\label{eq:sample_crps}
\end{equation}
which is implemented, for example, in the \texttt{scoringRules} package in \texttt{R}. 

For a confidence level \(\alpha \in (0,1)\), the \((1-\alpha)\times 100\%\) central prediction interval, defined by the \(\alpha/2\) and \(1-\alpha/2\) quantiles, allows assessment of sharpness, measured by the average width, and calibration, quantified by the proportion of observations contained in the interval (coverage). For a \(K\)-member ensemble, setting \(\alpha = 2/(K+1)\) aligns the nominal coverage with the empirical coverage \((K-1)/(K+1)\times 100\%\) of the raw ensemble \citep[see e.g.][]{gneiting2007strictly}.

%If $\mathbf{X}$ and $\mathbf{X}'$ are independent $D$-dimensional random vectors with distribution $F$, 
The generalization of Equation \eqref{eq:sample_crps} leads to 
the sample energy score \citep[ES;][]{gneiting2007strictly}. For an ensemble forecast $\mathbf{f}_k = \big(f_k^{(1)}, f_k^{(2)}, \dots, f_k^{(D)}\big)^\top$, $k = 1, \dots, K$, and the corresponding observation vector $\mathbf{y} = \big(y^{(1)}, y^{(2)}, \dots, y^{(D)}\big)^\top$, the sample energy score is given by
\begin{equation*}
    \mathrm{ES}(\hat{F}_K, \mathbf{y}) = \frac{1}{K} \sum_{k=1}^{K} \|\mathbf{f}_j - \mathbf{y}\| - \frac{1}{2K^2} \sum_{k=1}^{K} \sum_{\ell=1}^{K} \|\mathbf{f}_k - \mathbf{f}_\ell\|,
\end{equation*}
where $\|\cdot\|$ denotes the Euclidean norm.

In addition to the ES, the multivariate performance of ensemble forecasts is also assessed using the variogram score of order $p$ \citep[VS$_p$][]{scheuerer2015variogram}, which is particularly sensitive to errors in the correlation structure of the forecasts. The variogram score of order $p$ is defined as
\begin{equation*}
    \mathrm{VS}_p(F_K, \mathbf{y}) = \sum_{i=1}^D \sum_{j=1}^D \omega_{ij} \left( |y^{(i)} - y^{(j)}|^p - \frac{1}{K} \sum_{k=1}^K |f_k^{(i)} - f_k^{(j)}|^p \right)^2,
\end{equation*}
where $\omega_{ij} \geq 0$ are weights that quantify the relative importance of each coordinate pair $(i,j)$.

Common choices for the order $p$ are $0.5$ and $1$. In our study, we use $p=0.5$ and denote the score simply as VS.

Graphical tools for assessing the reliability of multivariate forecasts, such as multivariate rank histograms, are also well established. These histograms are based on pre-ranks, which condense a multivariate quantity into a single value. Since interpreting multivariate histograms can be challenging, it is advisable to employ multiple pre-ranking methods concurrently to obtain a comprehensive view of forecast miscalibration \citep{allen2024assessing}. Following this strategy, we consider average-, band depth-, energy score-, and dependence-based histograms in our analysis. The average rank is computed as the mean of the ranks of observations within the ensembles. Its histogram offers a diagnostic assessment of the forecast distribution: underdispersion is reflected by a $\cup$-shaped histogram, overdispersion by a $\cap$-shaped histogram, and bias typically results in a triangular pattern, while the band depth rank of each observation reflects its centrality relative to the ensemble of forecasts \cite{thorarinsdottir2016assessing}.  Whereas the energy score and dependence histograms build on the more recent concept of proper score-based pre-ranks introduced by \cite{knuppel2022}.

In addition, the reliability index (RI) provides a quantitative measure of the uniformity of a rank histogram, thereby indicating the degree to which the ensemble is properly calibrated. For a histogram with \(K+1\) bins and predicted frequencies \(\hat{p}_k\) in each bin, the reliability index is defined as
\[
\mathrm{RI} = \sum_{k=1}^{K+1} \left| \hat{p}_k - \frac{1}{K+1} \right|,
\]
where smaller values of \(\mathrm{RI}\) indicate a more uniform histogram.

The statistical significance of differences in verification scores is assessed using the Die\-bold–Mariano (DM) test \citep{diebold2002comparing}, which accounts for temporal dependencies in forecast errors. Given a scoring rule and two competing forecasts, \(F\) and \(F_{\mathrm{ref}}\), the DM test statistic is defined as  
\begin{equation*}
    t = \frac{\bar{d}}{\hat{\sigma}_{\bar{d}} / \sqrt{n}},
    \label{eq:dm_stat}
\end{equation*}
where \(n\) denotes the number of forecast cases in the test set, \(\bar{d}\) is the mean difference in scores over the test set between forecasts \(F\) and \(F_{\mathrm{ref}}\), and \(\hat{\sigma}_{\bar{d}}\) \ is a consistent estimator of the asymptotic standard deviation of the individual score differences. Under standard regularity conditions, the DM statistic asymptotically follows a standard normal distribution under the null hypothesis of equal predictive accuracy. Negative values of the DM statistic indicate better predictive performance of \(F\), while positive values favor \(F_{\mathrm{ref}}\). To account for multiple testing, the Benjamini–Hochberg correction \citep{benjamini1995controlling} is applied.

In the case studies presented in Section \ref{sec5}, improvements of forecasts relative to a reference forecast \(F_{\mathrm{ref}}\) are quantified using skill scores derived from a verification metric \(S\) \citep{gneiting2007strictly}. The skill score is defined as
\[
\text{Skill Score} = 1 - \frac{\overline{S}_F}{\overline{S}_{F_{\mathrm{ref}}}},
\]
where \(\overline{S}_F\) and \(\overline{S}_{F_{\mathrm{ref}}}\) denote the mean values of the score \(S\) over all forecast cases in the verification period for the forecast \(F\) and the reference \(F_{\mathrm{ref}}\), respectively. When \(S\) corresponds to the CRPS, this skill score is commonly referred to as the continuous ranked probability skill score (CRPSS). In addition to the CRPSS, we evaluate forecasts using the energy skill score (ESS) and the variogram skill score (VSS), with higher values indicating better performance.

\begin{figure}[t!]
    \centering
    \includegraphics[width=.8\linewidth]{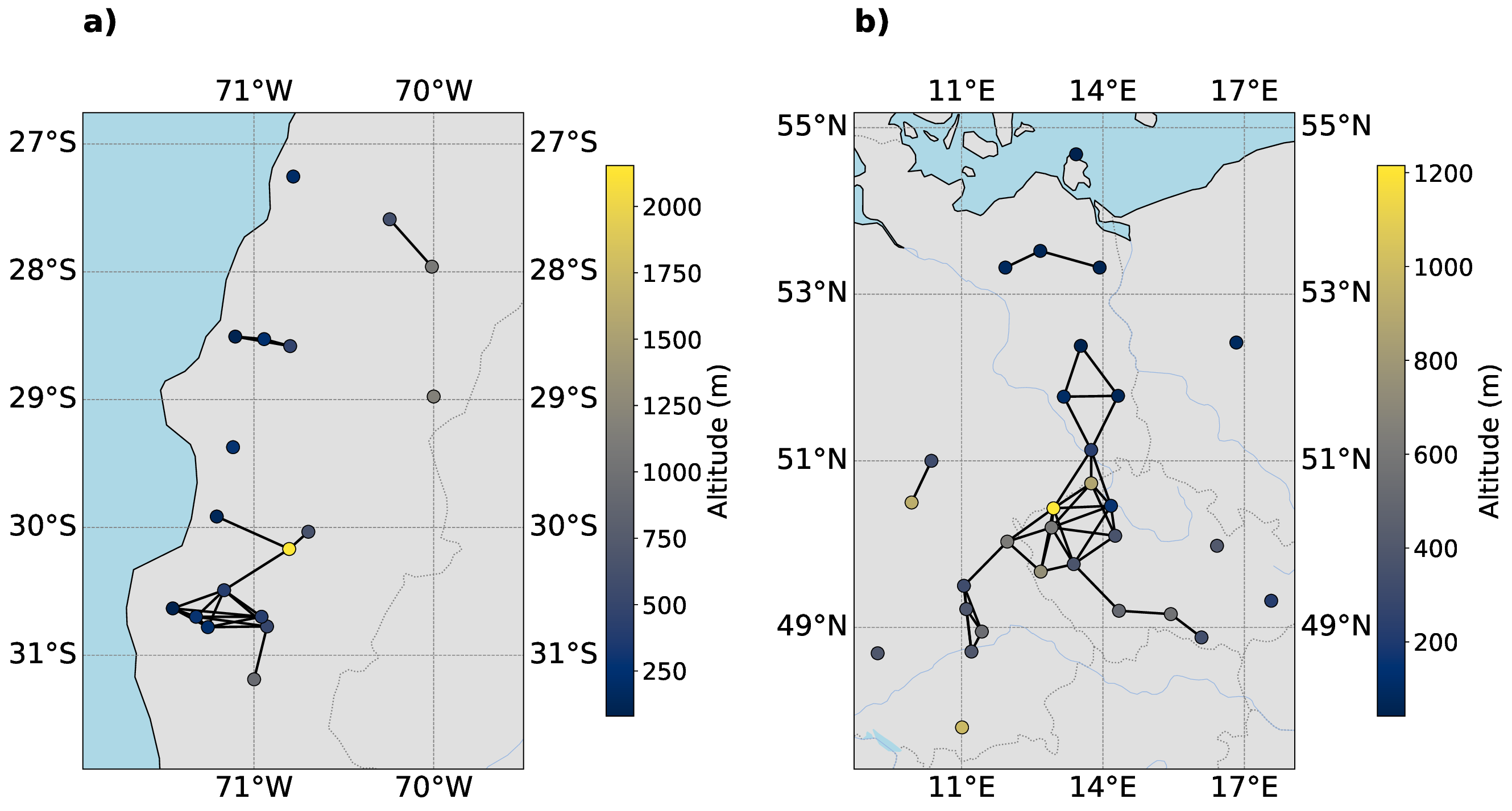}
    \caption{Locations of solar irradiance observation stations in northern Chile (a) and visibility observation stations across Central Europe (b). Black lines indicate edges used to construct the input graph for the GraphSAGE models.}
    \label{fig:locations_with_graphs}
\end{figure}

\section{Implementation details}

For the GNN model considered here, the spatial structure of the observational network is represented as an undirected graph 
\(\mathcal{G} = (\mathcal{V}, \mathcal{E})\), where the node set 
\(\mathcal{V} = \{1, 2, \ldots, D\}\) corresponds to the observation stations, and the edges 
\(\mathcal{E}\) capture spatial relationships. An edge is established between two nodes \(d\) and \(d'\) 
if the distance between the corresponding stations is below a threshold \(r\), which is selected to minimize the CRPS. 
The graph yielding the best performance for both datasets is shown in Figure \ref{fig:locations_with_graphs}. 
The threshold is set to \(r = 50\,\mathrm{km}\) for solar irradiance and \(r = 100\,\mathrm{km}\) for visibility.

In both case studies, we use the GraphSAGE architecture \citep{hamilton2017inductive} with the mean aggregator, differing only in minor hyperparameter settings. 
For solar irradiance, the network consists of a single hidden \texttt{SAGEConv} layer mapping the input features to a hidden representation, followed by batch normalization, ReLU activation, and dropout. 
For visibility, the network uses two hidden \texttt{SAGEConv} layers with the same setup for each layer. 
In both cases, a final \texttt{SAGEConv} layer outputs the forecast-specific representation without an activation function. 
In principle, a non-negativity constraint could be enforced through activation functions, which would be relevant for both of the meteorological variables considered. 
However, to ensure architectural consistency with the MLP of \cite{bmcdsznl25}, the non-negativity requirement was implemented within the cost functions instead.

Training was conducted for up to 500 epochs, employing early stopping based on validation loss with a patience of 15 epochs for solar irradiance and 10 epochs for visibility.
The hyperparameters of the GNN models are summarized in Table \ref{tab:gnn_hyperparameters_all}.

For solar irradiance, each observation station is characterized by features including the ensemble mean, the relative frequency of zero irradiance values, and the ensemble variance for the given lead time and day. Additionally, the station’s geographical coordinates (latitude, longitude), elevation, and the forecast lead time are included to distinguish different forecast hours within the day. This feature set is fully consistent with that used in \cite{bmcdsznl25}.

For visibility, each observation station is represented by a feature set that includes the control forecast, the mean and standard deviation of the 50 exchangeable ensemble members, and the proportions of forecasts predicting visibility below 5000 m, between 5000 and 30,000 m, and 30,000--70,000 m. Additionally, visibility point forecasts from the Copernicus Atmosphere Monitoring Service (CAMS) are incorporated, along with annual base functions to capture seasonal and temporal effects, defined as
\begin{equation*}
    \beta_1(d):=\sin \big(2\pi d/365 \big) \qquad \text{and} \qquad  \beta_2(d):=\cos \big(2\pi d/365 \big),
  \end{equation*}
where \ $d$ \ denotes the day of the year. Furthermore, the station’s latitude, longitude, and elevation are also taken into account.

As with the solar irradiance data, this feature set largely follows \cite{lakatos2024enhancing}; however, for the GNN, the geographical coordinates are explicitly included to enable graph construction and capture location-dependent information. Lead time is also included to account for temporal variations.

\begin{table}[t]
\centering
\label{tab:gnn_hyperparameters_all}
\begin{tabular}{lcc}
\hline
\textbf{Parameter}               & \textbf{Solar Irradiance}         & \textbf{Visibility}                \\ \hline
Number of nodes/stations \(D\)  & 18                               & 30                               \\
Edge definition                 & Distance \(< 50\,\mathrm{km}\)    & Distance \(< 100\,\mathrm{km}\)   \\
Aggregator                     & Mean aggregator                  & Mean aggregator                  \\
Number of hidden \texttt{SAGEConv} layers & 1                                & 2                                \\
Number of hidden units          & 1024                             & 64                               \\
Output dimension \(K\)              & 8                                & 51                                \\
Dropout                       & 0.2                              & 0.2                              \\
Batch size                    & 64                               & 64                               \\
Learning rate                 & 0.03                             & 0.03                             \\ 
Validation set size                 & 0.3                             & 0.3                             \\ 
Number of epochs                 & 500                             & 500                           \\ 
\hline
\end{tabular}
\caption{Hyperparameter settings of the GNN models for solar irradiance and visibility forecasting}
\end{table}

For both case studies, a single model was initialized to process all lead times jointly. To assess uncertainty, each model was trained and evaluated ten times during testing, with performance measured by the average verification scores across these runs.  The models generally exhibited stable behavior, with only minor performance differences observed between iterations.

The hyperparameters listed in Table \ref{tab:gnn_hyperparameters_all} were first found by minimizing the CRPS. Subsequently, multivariate loss functions were introduced to enable the GNN to more effectively capture dependencies between stations. Specifically, apart from the variant trained solely by minimizing the ES, we employed a weighted combination of the ES and the VS:
\[
\mathcal{L} = w_1 \cdot \text{ES} + w_2  \cdot \text{VS},
\]
where \(\alpha\) and \(\beta\) are weighting coefficients. In our experiments, we set \(w_1 \in [0, 1]\) and \(w_2 = 1 - w_1\), for simplicity and interpretability, although in principle other values (e.g. \(w_1 > 1\)) could be used. Due to the large difference in scale between the two terms--the VS was found empirically to be several orders of magnitude larger than the ES--the VS component was normalized by an appropriate factor. Specifically, this factor was computed as the ratio of the mean ES to the mean VS over the raw ensemble, which was consistent with the ratios observed in the training batches. This normalization allowed us to select a weighting within the \([0, 1]\)  interval that balances the contributions of both terms, ensuring that neither dominates the loss and that optimization proceeds stably. The GNN models were implemented in Python using the \texttt{PyTorch Geometric} library \citep{fey2019fast}.
\subsection{Benchmark methods}
\label{sec4.1}

For the post-processing of solar irradiance forecasts, we employed the MLP model previously used by \citet{bmcdsznl25} on the same dataset, adopting the same hyperparameters as in their study. Specifically, a 25-day rolling training period was used, with a batch size of 1200 and a learning rate of 0.01. The input features were identical to those summarized earlier, and the network architecture consisted of two hidden layers with 255 neurons each. Early stopping was implemented with a patience of 5 epochs, and the test set comprised 20\% of the data.

\begin{figure}[t!]
    \centering
    \includegraphics[width=1\textwidth]{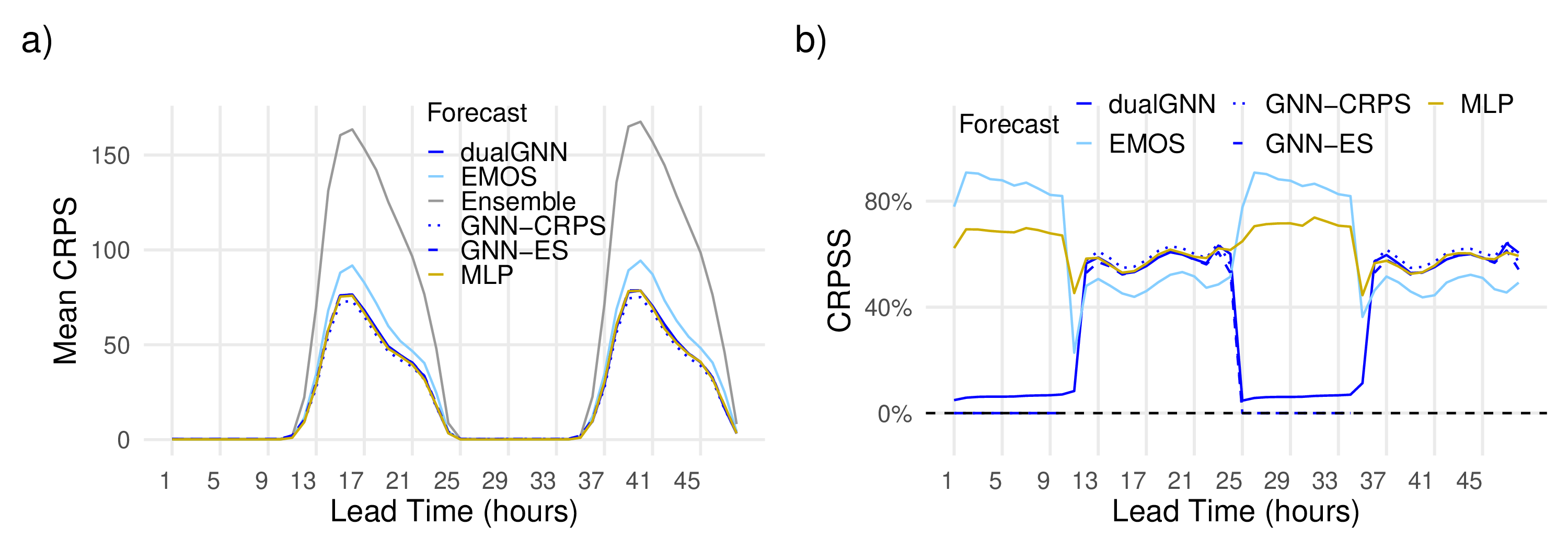}
   \caption{(a) Mean CRPS of raw, EMOS-, and GNN-based post-processed solar irradiance forecasts; (b) CRPSS of post-processed forecasts relative to the raw ensemble. Both panels are shown as functions of the lead time.}
    \label{fig:crps_crpss}
\end{figure}

\begin{table}[t!]
\centering
\begin{tabular}{lccccc}
\toprule
 & EMOS & MLP & dualGNN & GNN-ES & GNN-CRPS \\
\midrule
 & 51.62\% & 42.57\%  & 42.85\% & 43.23\% & 40.79\% \\
\bottomrule
\end{tabular}
    \caption{Overall mean CRPS of post-processed solar irradiance forecasts as proportion of the mean
    CRPS of the raw ensemble for observed solar irradiance not less than 7.5 $W/m^2$.}
\label{crps_tr}
\end{table}
For the EMOS model, we applied the semi-local approach of \cite{lerch2017similarity}, briefly described in Section \ref{sec3.1}. In this case, the 18 stations were grouped into three clusters, each containing at least two stations. The rolling training period was set to 80 days based on an extensive hyperparameter search.

Finally, the reference forecasts used as the basis for the visibility dualGNN were generated by the POLR classifier described in Section \ref{sec3.2}. To ensure consistency with the dualGNN outputs, a sample of size 51 was drawn from the predictive PMFs produced by these classifiers. These models were trained over a 350-day period and used the same set of features as in \cite{lakatos2024enhancing}.

\section{Results}
\label{sec5}

In the following, we present the results of post-processing applied to forecasts of solar irradiance and visibility. While our main focus is on improving the representation of spatial dependencies between locations, we also provide a brief assessment of the marginal forecast accuracy for each variable, which enables evaluation of both individual performance and the effectiveness in capturing spatial correlations across observation sites.

\subsection{Univariate performance of solar irradiance forecasts}
\label{sec:uv_results_rad}

 \begin{figure}[t!]
    \centering
    \includegraphics[width=1\textwidth]{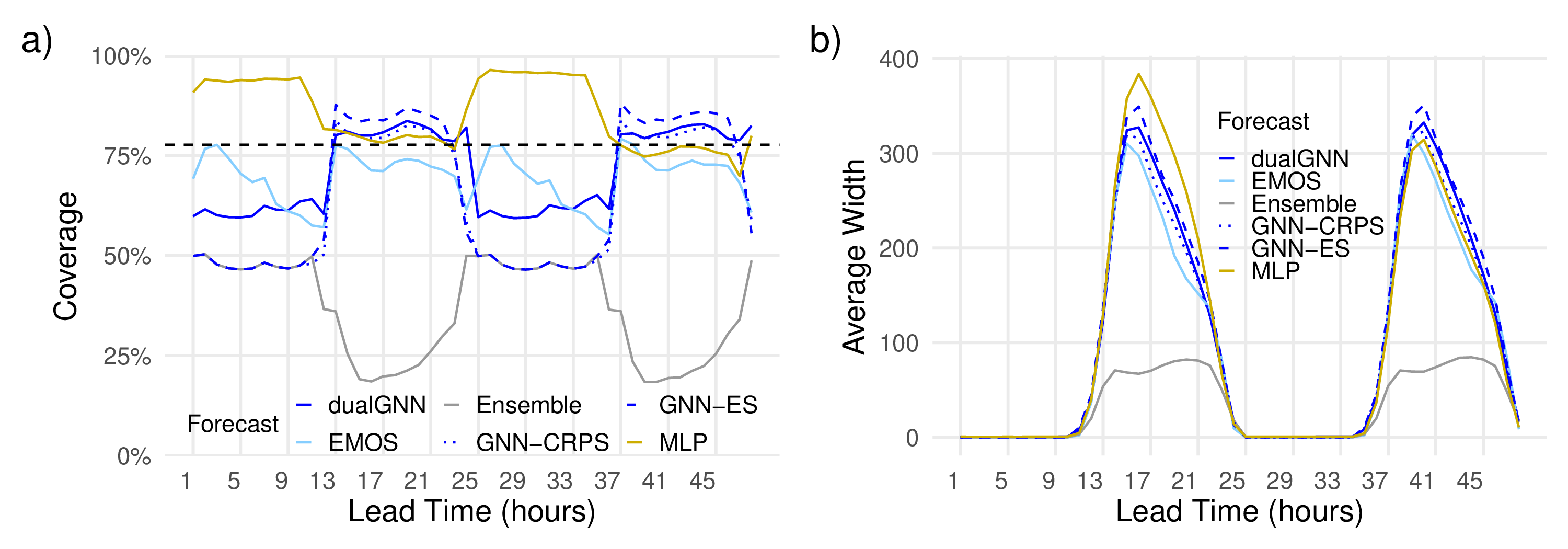}
   \caption{(a) Coverage and (b) average width of the 77.78\% central prediction intervals for post-processed and raw solar irradiance forecasts across lead times. The dashed line in (a) indicates the nominal coverage.}
    \label{fig:cov_rad}
\end{figure}

As discussed in Section \ref{sec4.1}, for the post-processing of solar irradiance forecasts, following \cite{bmcdsznl25}, a semi-local EMOS model is utilizing an 80-day training period, whereas 25 days proved optimal for the MLP. Although GNNs typically require larger datasets for effective parameter estimation, a 30-day training period yielded the best performance in this case, likely due to maintaining relevance to the current atmospheric conditions. Both the MLP and the dualGNN generate 8-member ensemble forecasts to ensure comparability with the raw WRF forecasts. For the same purpose, we also consider eight equidistant quantiles extracted from the censored normal EMOS predictive distributions, with these samples hereafter referred to simply as EMOS. For a more comprehensive evaluation, we also consider GNN variants optimizing only ES or CRPS—denoted GNN-ES and GNN-CRPS—alongside the dualGNN. Model hyperparameters were selected by CRPS minimization, and the resulting architectures were subsequently tested using ES and the dual loss, allowing assessment of the contribution of each loss function to overall model performance. We tested different ES–VS weightings in the dual loss, assessing each by the proportion of cases that were statistically significantly better than the calibrated references according to the DM test. Excessively high ES weights reduced the proportion of better VS cases, while high VS weights degraded ES performance and CRPS. The 0.9–0.1 ES–VS weighting achieved the best balance, simultaneously maximizing the proportions of significantly better ES and VS cases and maintaining reasonable mean CRPS. Notably, adding the VS component did not compromise ES performance but substantially improved VS and also ES values compared to GNN models trained with only ES or CRPS loss.

\begin{figure}[t!]
    \centering
    \includegraphics[width=1\linewidth]{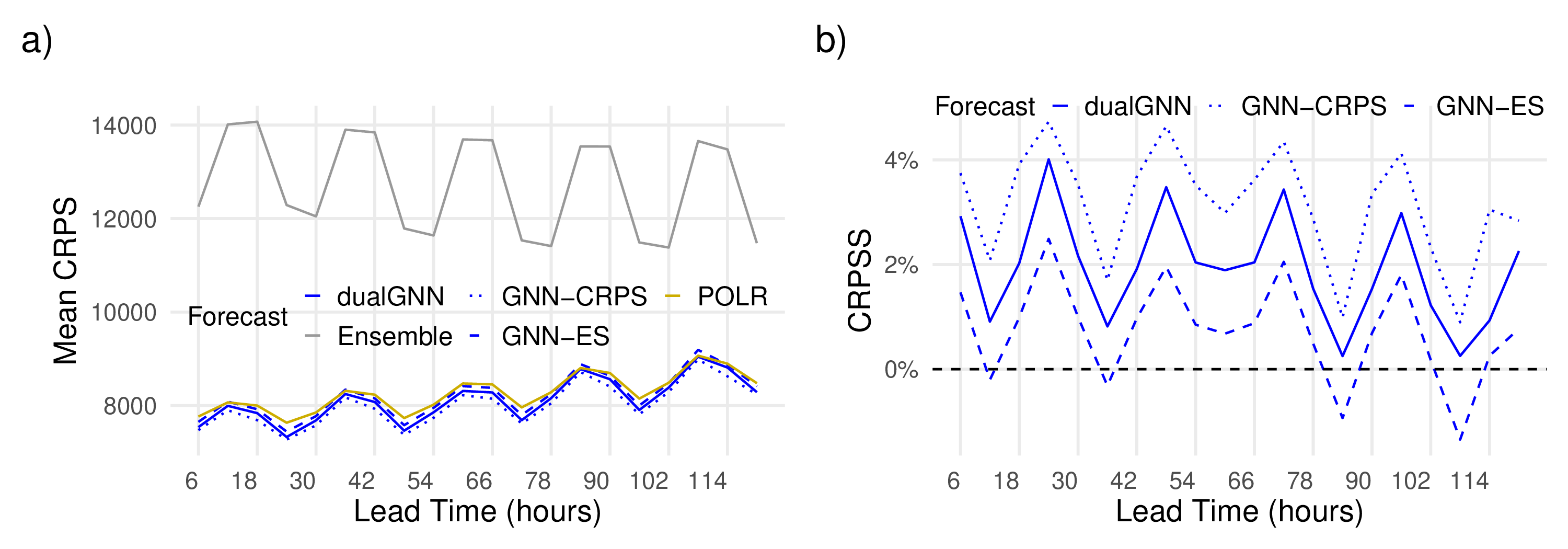}
    \caption{(a) Mean CRPS of raw, POLR-, and GNN-based post-processed visibility forecasts; (b) CRPSS of post-processed forecasts relative to the POLR predictions. Both panels are shown as functions of the lead time.}
    \label{fig:crps_crss_vis}
\end{figure}

\begin{table}[t!]
\centering
\begin{tabular}{lcccc}
\toprule
 & POLR & dualGNN & GNN-CRPS & GNN-ES \\
\midrule
 & 64.92\%  & 63.69\%  & 62.9 \% & 64.46 \%\\
\bottomrule
\end{tabular}
\caption{Overall mean CRPS of post-processed visibility forecasts as proportion of the mean
CRPS of the raw ensemble.}
\label{crps_props_vis}
\end{table}

Figure \ref{fig:crps_crpss} displays the mean CRPS and the corresponding CRPSS values relative to the raw ensemble forecasts. As of Figure \ref{fig:crps_crpss}a, all post-processed forecasts substantially improve upon the raw ensemble during daylight hours (1200–2400 UTC, corresponding to lead times of 12–24 h and 36–48 h), which are particularly relevant for solar irradiance applications. During this period, the EMOS model provides the least improvement, while the GNN and MLP models perform nearly identically, both clearly outperforming the raw forecasts and EMOS. Figure \ref{fig:crps_crpss}b highlights performance differences during nighttime hours, where the advantage of the more complex GNN models diminishes, and the EMOS model shows the largest improvement over the raw ensemble. In contrast, during daylight hours the GNN models and the MLP consistently provide the largest gains, with an average improvement of around 52\%. The ordering of the post-processed forecasts remains the same across both days. This overall ranking becomes even clearer when examining Table \ref{crps_tr}, which reports the mean CRPS of the post-processed solar irradiance forecasts as a proportion of the mean CRPS of the raw ensemble for cases with observed irradiance above $7.5$ $W/m^2$. This threshold, recommended by forecasters at the Hungarian Meteorological Service, reflects the fact that predictive performance is of practical importance primarily when solar irradiance reaches levels relevant for photovoltaic energy production \citep{bb24, bmcdsznl25}. Based on this metric, the GNN-CRPS model yields the largest improvement, followed by the MLP, and closely by the dualGNN.

Figure \ref{fig:cov_rad} displays the coverage and average width of the nominal 77.78\% central prediction intervals for post-processed and raw solar irradiance forecasts across lead times. Here, the coverage of the entire ensemble range is considered. Based on Figure \ref{fig:cov_rad}a, all post-processed forecasts exhibit better calibration than the raw ensemble forecasts. Considering the mean absolute deviation from the nominal coverage across all observations, the dualGNN is closest to the nominal level, followed closely by the MLP.  When focusing only on observations during 1200–2400 UTC, the MLP and GNN-CRPS models show equally good calibration, with the dualGNN slightly behind. As generally expected, improved calibration is associated with wider prediction intervals, with the MLP producing the widest intervals for the first 24 hours, followed by the GNN models and EMOS. In the second 24 hours the order reverses, and the GNN models produce slightly wider intervals than the MLP. Overall, the GNN intervals remain relatively consistent across both periods, while the MLP intervals narrow slightly in the second 24 hours, approaching those of EMOS.

\begin{figure}[t!]
    \centering
    \includegraphics[width=1\linewidth]{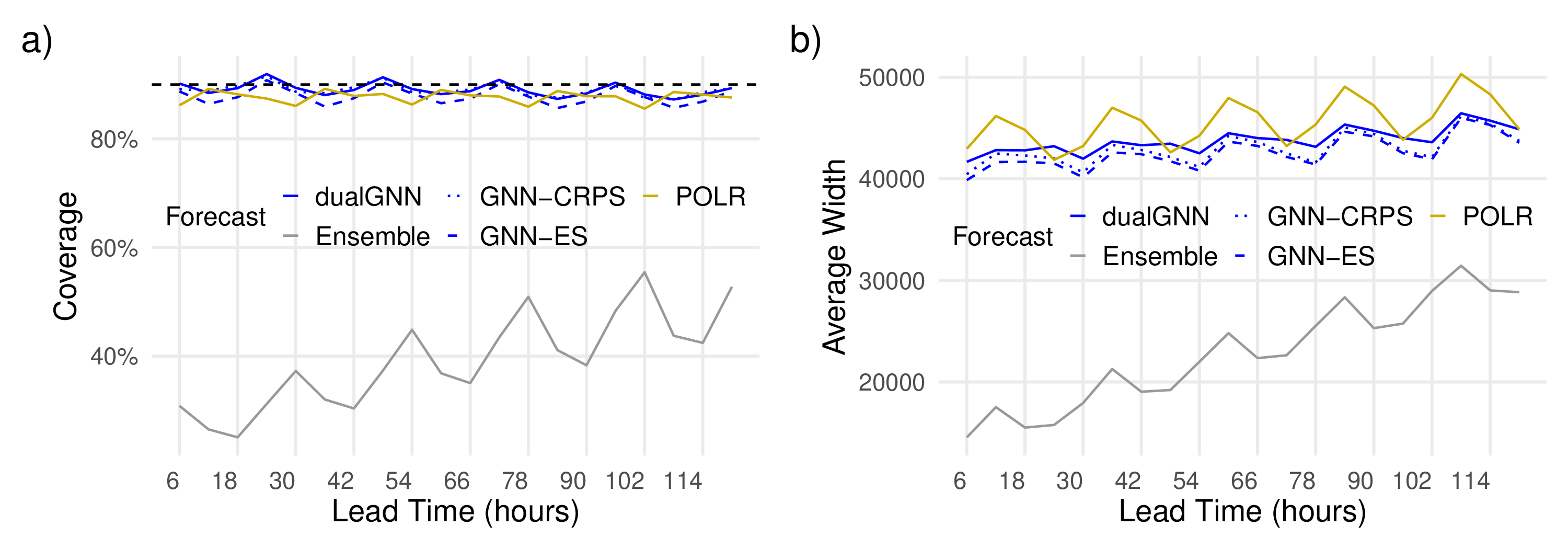}
    \caption{Coverage (a) and average widths (b) of 90\% central prediction intervals of raw and post-processed visibility forecasts as functions of the lead time. In panel (a) the ideal coverage is indicated by the horizontal dashed line.}
    \label{fig:cov_avgWidth_vis}
\end{figure}

\subsection{Univariate performance of visibility forecasts}
\label{sec:uv_results_vis}

For visibility forecast post-processing, we use the POLR classifier introduced in Section \ref{sec3.2} as a reference alongside the GNN models. POLR provides a strong baseline, as it is specifically designed for ordered data such as the visibility observations considered here. Similar to the solar irradiance predictions, the GNN models generate forecasts that are both calibrated and spatially consistent, using the same discrete observational categories as POLR. Unlike POLR, which requires a maximum to be imposed on values derived from the probability mass functions (PMFs), the GNN models can produce values of any magnitude, unconstrained by the historical observation range. The POLR model employs a 350-day rolling training period; however, this was insufficient for the GNN models, which were trained using 530 days of data. Consistent with the approach in Section \ref{sec:uv_results_rad}, we evaluate three variants of the GNN models: the dualGNN, which optimizes a composite loss function, and the GNN-CRPS and GNN-ES models, which minimize the CRPS and the ES, respectively. Following a similar approach to the EMOS models described in Section \ref{sec:uv_results_rad}, the POLR forecast PMFs are represented by drawing 51 equidistant quantiles, enabling direct comparison with both the raw ensemble forecasts and the GNN-generated samples. Similar to the approach discussed in the previous section, the dualGNN employs a loss function that combines the ES and VS components to achieve the most favorable trade-off between multivariate and univariate scores. In the present analysis, the model minimizes ES with a weight of 0.3 and VS with a weight of 0.7.  As with the radiation data, increasing the weight of ES improves ES performance, while placing greater weight on VS enhances VS performance. The chosen weighting scheme not only provides a balanced compromise but also yields the lowest mean CRPS among all dualGNN variants tested with different weightings.

\begin{figure}[t!]
    \centering
    \includegraphics[width=1\textwidth]{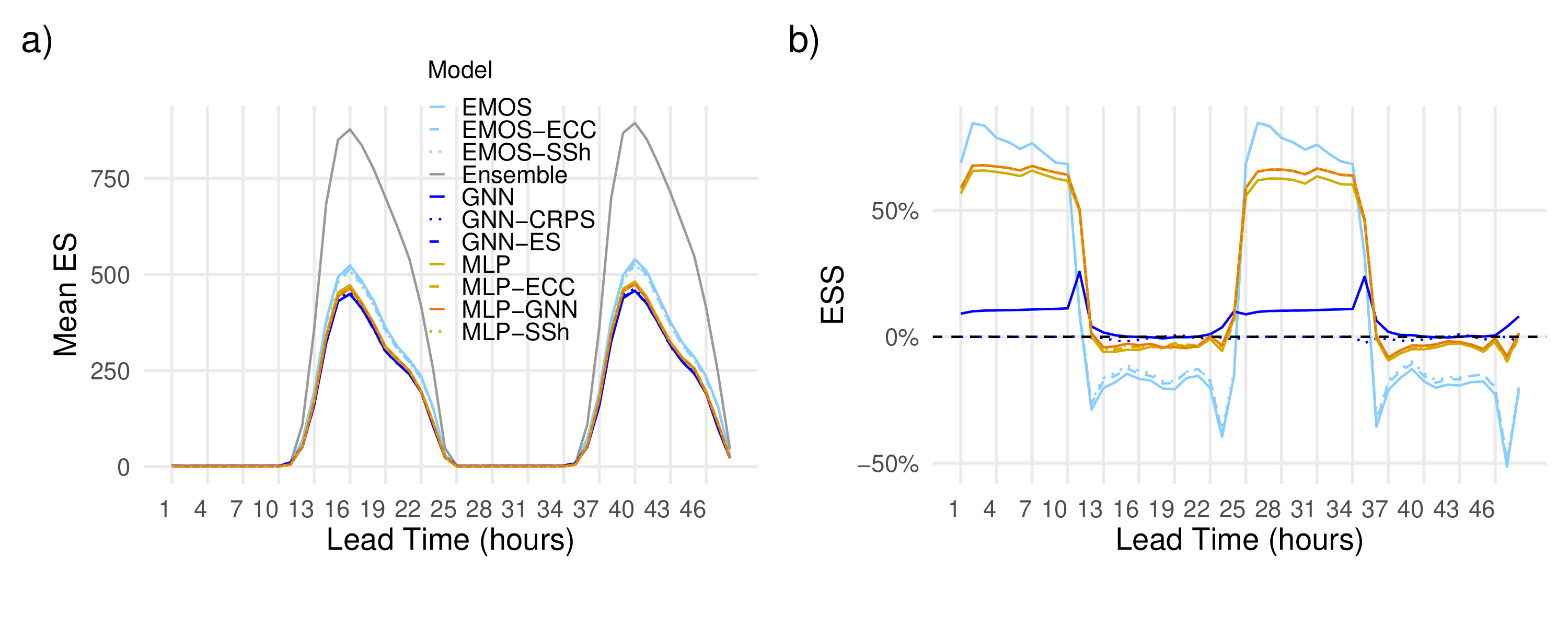}
    \caption{(a) Mean ES of raw, independently post-processed (EMOS), and multivariate solar irradiance forecasts; (b) ESS of post-processed forecasts relative to GNN-ES.
Both shown as functions of the lead time.}
    \label{fig:es_ess_rad}
\end{figure}

Figure \ref{fig:crps_crss_vis} shows the mean CRPS of raw, POLR-, and GNN-based post-processed visibility forecasts, and the CRPSS of the post-processed forecasts relative to the POLR predictions, both as functions of lead time. As in the previous case study, Figure \ref{fig:crps_crss_vis}a indicates that all post-processed models outperform the raw ensemble forecasts; however, this advantage decreases with increasing lead time for all models. Differences in model performance are more clearly visible in Figure \ref{fig:crps_crss_vis}b, which presents the skill scores, with respect to the reference POLR model. Here, the GNN-CRPS model provides the largest improvement relative to POLR, followed by the dualGNN, while the GNN-ES model also shows positive gains over POLR for almost all lead times. The ranking of the GNN models is further confirmed by Table \ref{crps_props_vis}, which presents the overall mean CRPS of the post-processed visibility forecasts as a proportion of the mean CRPS of the raw ensemble, with the best-performing GNN-CRPS model showing a 2.02\% improvement over POLR and the dualGNN showing a 1.23\% improvement.

Finally, Figure \ref{fig:cov_avgWidth_vis} presents the coverage and average widths of the 90\% central prediction intervals for raw and post-processed visibility forecasts as functions of lead time. The GNN models exhibit a mean absolute deviation of approximately 1\% from the nominal 90\% coverage, compared to 2\% for POLR and 51\% for the raw ensemble. Moreover, temporal dependencies and the diurnal cycle are less pronounced in the coverage values of all post-processed models than in the raw ensemble. Figure \ref{fig:cov_avgWidth_vis}b illustrates that the raw ensemble produces the narrowest prediction intervals, whereas the GNN models have narrower average widths than POLR and are less sensitive to the diurnal cycle.

\begin{figure}[t!]
    \centering
    \includegraphics[width=1\textwidth]{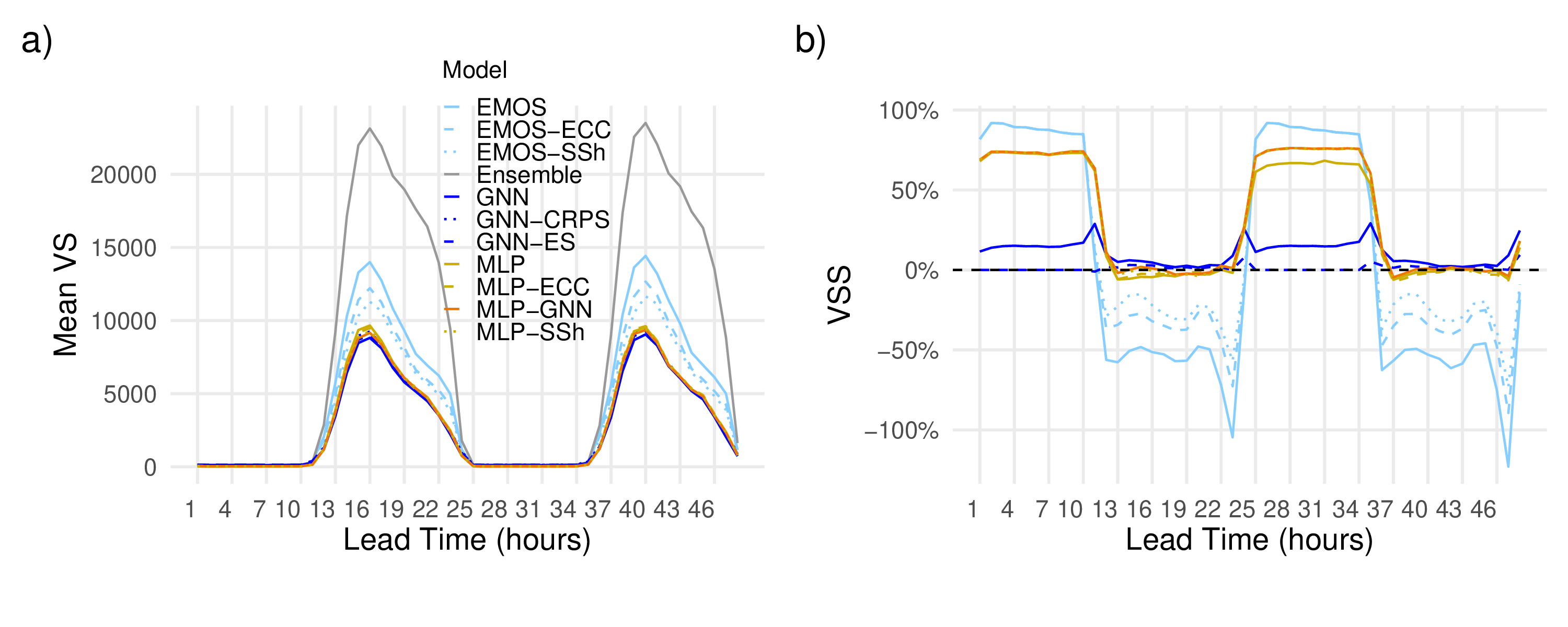}
    \caption{(a) Mean VS of raw, independently post-processed (EMOS), and multivariate solar irradiance forecasts; (b) VSS of post-processed forecasts relative to GNN-ES.
Both shown as functions of the lead time.}
    \label{fig:vs_vss_rad}
\end{figure}

\subsection{Multivariate performance of solar irradiance forecasts}
\label{sec:mv_results_rad}
In this section, the performance of the proposed GNN models is assessed in comparison with a range of empirical copula-based approaches. Specifically, when the EMOS and MLP samples introduced in Section \ref{sec:uv_results_rad} are reordered to match the rank structure of the raw WRF forecasts, the resulting methods are denoted EMOS-ECC and MLP-ECC, respectively. Alternatively, when the dependence template is derived from past observations within the training period, the approaches are referred to as EMOS-SSh and MLP-SSh. As shown by \cite{lakatos2023comparison} restricting the set of possible pool dates to the training period does not deteriorate forecast performance. Furthermore, for the current dataset, we include in the analysis a hybrid model that combines the MLP with the dualGNN by mapping the MLP-generated samples onto the rank order structure of the dualGNN forecasts for the same verification day and forecast horizon. Consequently, this hybrid model requires running the MLP and dualGNN in parallel. This reference model, hereafter referred to as MLP-GNN, enables us to evaluate whether the dualGNN rank structure provides additional information beyond that contained in the raw ensemble forecasts and historical observations. To provide a baseline, we also consider the EMOS and MLP samples under the assumption of independence, meaning that the calibrated samples are not rearranged according to any dependence template.

\begin{figure}[t!]
    \centering
    \includegraphics[width=1\textwidth]{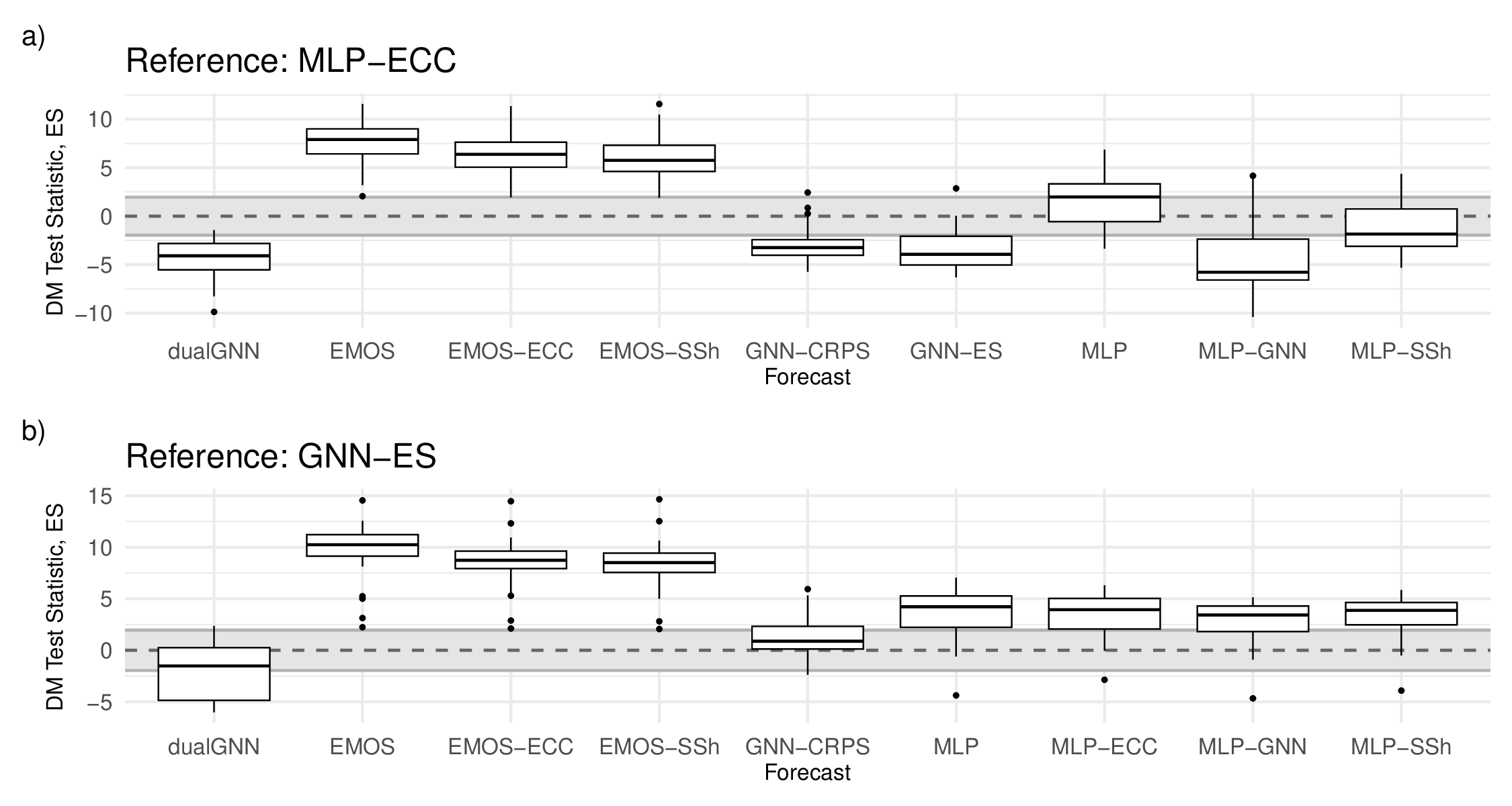}
    \ \caption{Boxplots of Diebold–Mariano (DM) test statistics investigating the significance of the difference in ES  (a) from the reference MLP-ECC and (b) GNN-ES methods as functions of the lead times 12–24 h and 36–48 h. Grey region indicates the acceptance region of the two-tailed DM test for equal predictive performance at a 5\% level of significance.}
    \label{fig:es_dm}
\end{figure}

Figure \ref{fig:es_ess_rad} displays the mean ES and the corresponding skill scores of the post-processed solar irradiance forecasts, both independent and multivariate, as functions of lead time, with the raw forecasts omitted from the skill score plot to better highlight the differences between the post-processed models. As shown in Figure \ref{fig:es_ess_rad}a, all post-processed models substantially improve the mean ES compared to the raw forecasts, and the ranking of the models remains consistent between the first and second 24 hours. Based on the mean ES, for cases when solar irradiance exceeds $7.5$  $W/m^2$, the models are ranked as follows: dualGNN, GNN-ES, GNN-CRPS, MLP-GNN, MLP-ECC, MLP-SSh, MLP, EMOS-SSh, EMOS-ECC, EMOS, and finally the raw ensemble forecasts. This ranking supports the conclusion that including the VS component in the loss function improves ES performance, indicating that incorporating VS into the training objective can enhance multivariate forecast evaluation based on ES. This is further shown in Figure \ref{fig:es_ess_rad}b, where the ES-minimizing GNN is used as a reference. Based on this comparison, the EMOS samples exhibit the weakest performance relative to the reference, with EMOS-ECC and EMOS-SSh showing nearly identical skill. The same pattern holds for the subsequent MLP variants and the CRPS-trained GNN. Although MLP-ECC and MLP-SSh outperform the standard MLP forecasts and the GNN-CRPS, their performance is largely comparable. These models are followed by MLP-GNN reordered according to the dualGNN rank structure, which is surpassed only by dualGNN, yielding an average improvement of 4\% over GNN-ES. The effect of including the VS in the loss function is most pronounced at the beginning and end of the day, i.e., during periods of low solar irradiance above 7.5 $W/m^2$, which correspond to the early morning and late afternoon hours. This suggests that the benefits of VS are particularly evident under low-irradiance conditions, where the spatial dependence captured by VS contributes most to improving forecast skill.

\begin{figure}[t!]
    \centering
    \includegraphics[width=1\textwidth]{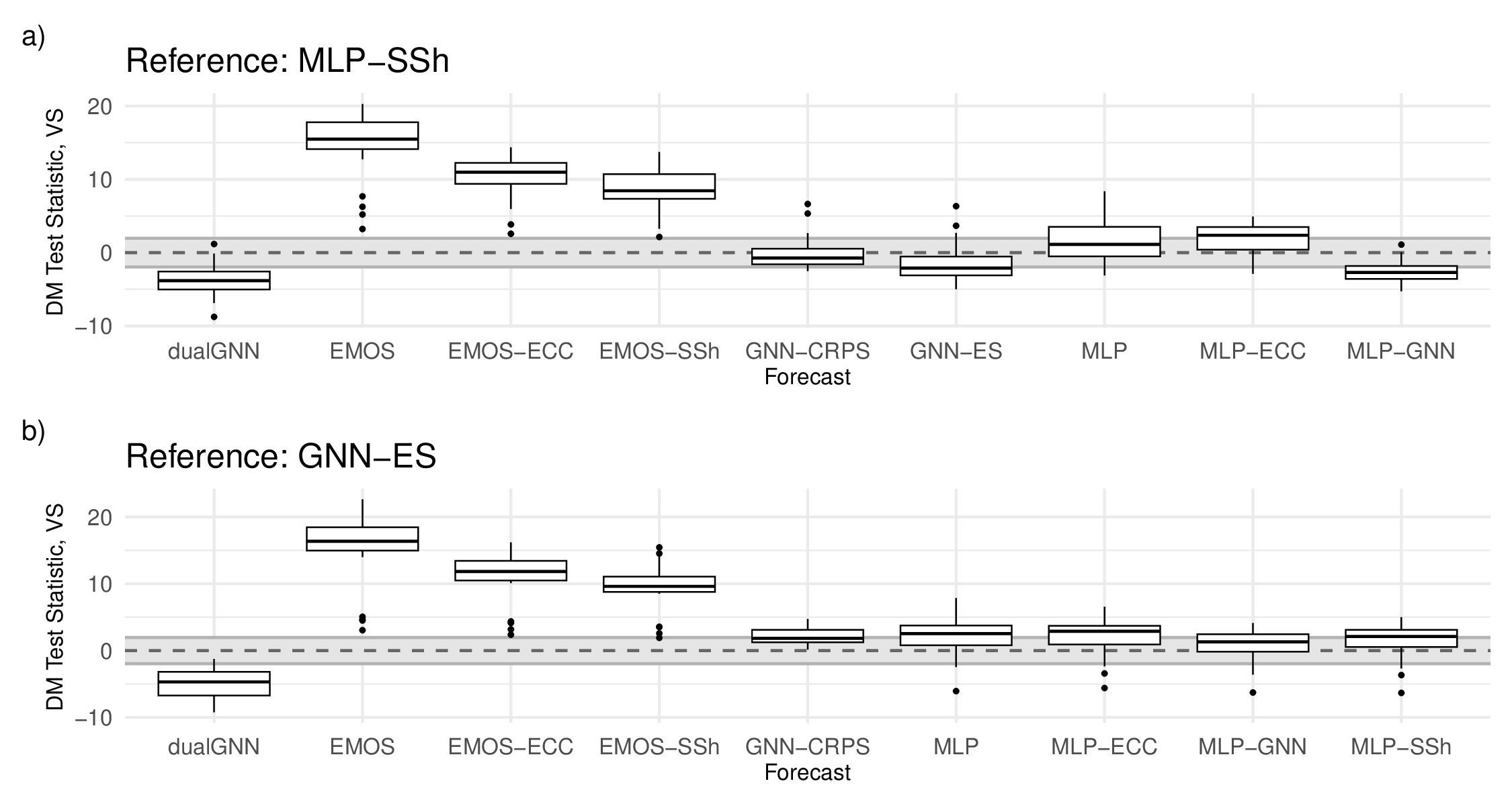}
    \ \caption{Boxplots of Diebold–Mariano (DM) test statistics investigating the significance of the difference in VS  (a) from the reference MLP-SSh and (b) GNN-ES methods as functions of the lead times 12–24 h and 36–48 h. Grey region indicates the acceptance region of the two-tailed DM test for equal predictive performance at a 5\% level of significance.}
    \label{fig:vs_dm}
\end{figure}

\begin{figure}[h!]
    \centering    
    \includegraphics[width=.83\linewidth]{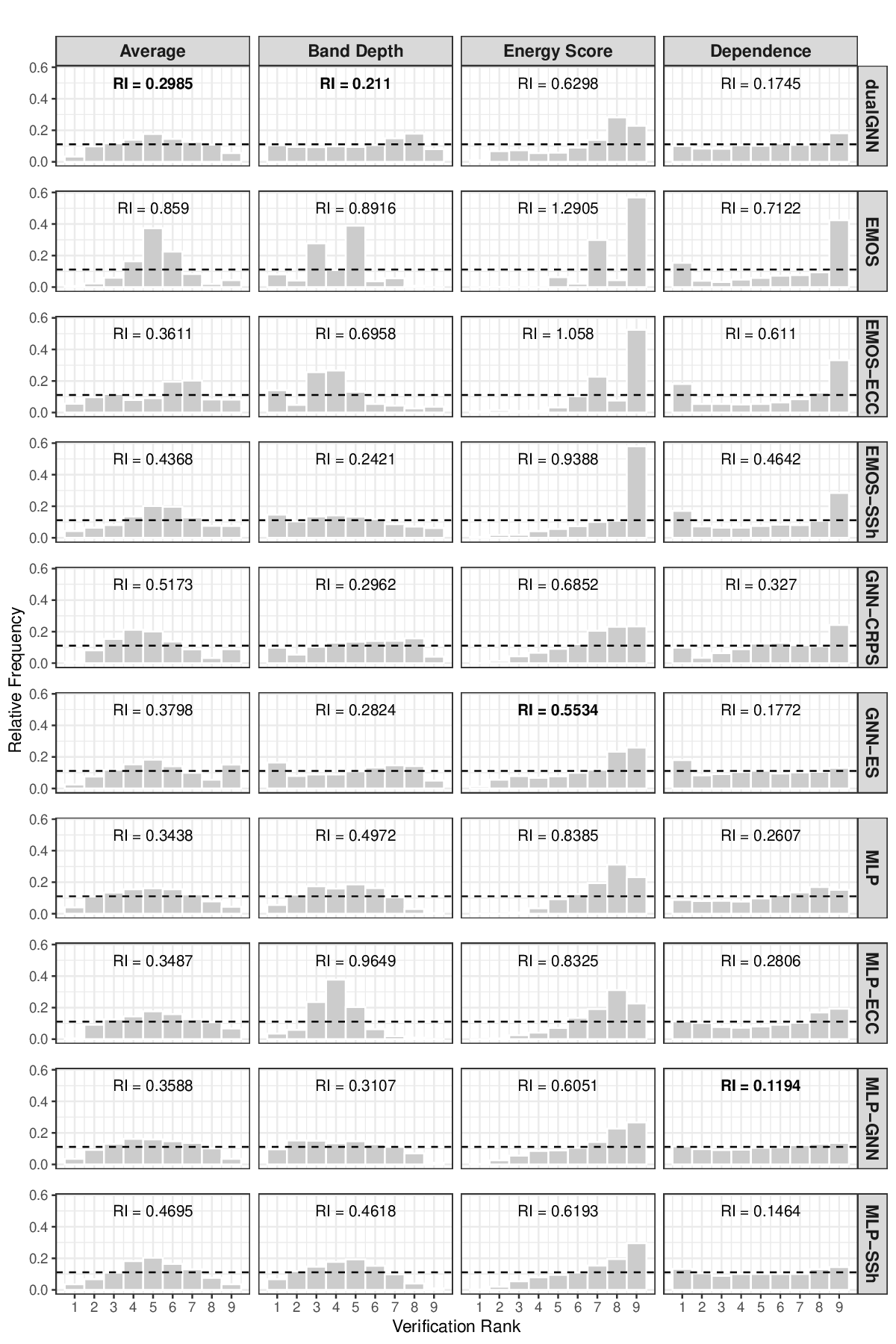}
    \caption{Rank histograms of post-processed solar irradiance  forecasts together with the
corresponding reliability indices for lead times 12–24 h and 36–48 h}
    \label{fig:mv_hists_rad}
\end{figure}

Figure \ref{fig:vs_vss_rad}, which shows the mean VS and corresponding skill scores of the raw, independently post-processed, and multivariate post-processed solar irradiance forecasts relative to GNN-ES, presents a very similar pattern. When forecasts are ranked according to the mean VS during periods in which the solar irradiance exceeds $7.5  W/m^2$, the models are ordered as follows: dualGNN, GNN-ES, GNN-CRPS, MLP-GNN, MLP-SSh, MLP, MLP-ECC, EMOS-SSh, EMOS-ECC, EMOS, and finally the raw ensemble. Once again, the independent EMOS model exhibits the weakest performance, followed by EMOS-ECC and EMOS-SSh. The ranking of the EMOS variants is therefore identical to that observed in Figure \ref{fig:es_ess_rad}a; however, the differences between models are more pronounced. Similarly to Figure \ref{fig:es_ess_rad}a, the model ranking is consistent across the two consecutive days. Differences among the GNN and MLP models are more clearly visible in Figure \ref{fig:vs_vss_rad}b, where GNN-ES is again used as the reference. Relative to this benchmark, during periods with solar irradiance above $7.5$  $W/m^2$, the EMOS model exhibits an average deficit of 65.82\%, followed by EMOS-ECC with 43.32\% and EMOS-SSh with 33.52\%. In contrast, MLP and GNN-CRPS models show substantial improvements over the EMOS variants, with skill scores of -2.2\%, -0.9\%, and -0.8\% for GNN-CRPS, MLP, and MLP-ECC, respectively. The MLP-SSh achieves a 0.1\% improvement relative to GNN-ES, effectively matching the reference model's performance. The model reordered according to the GNN rank structure shows a 1.3\% advantage, while dualGNN achieves an average improvement of 5.2\% over GNN-ES.

Figure \ref{fig:es_dm} presents boxplots of DM test statistics evaluating the significance of differences in ES relative to the strongest GNN-independent reference, MLP-ECC, and the GNN-ES method as functions of lead time. The grey shaded area indicates the acceptance region of the two-tailed DM test for equal predictive performance at a 5\% significance level, and the Benjamini–Hochberg procedure was applied to control the false discovery rate. Negative DM values correspond to better predictive skill compared to the reference.

As shown in Figure \ref{fig:es_dm}a, dualGNN, MLP-GNN, GNN-ES, and GNN-CRPS exhibit the greatest advantages, significantly outperforming the GNN-independent reference. Specifically, dualGNN performs better than the reference in 88.46\% of cases and is the only model that is never outperformed. For the other three models, this proportion is 76.92\%. Following these are MLP-SSh and MLP, which achieve improvements in 46.15\% and 15.38\% of cases, respectively. All EMOS variants, however, are outperformed by MLP-ECC in 96.15–100\% of cases.

\begin{figure}
    \centering    \includegraphics[width=1\linewidth]{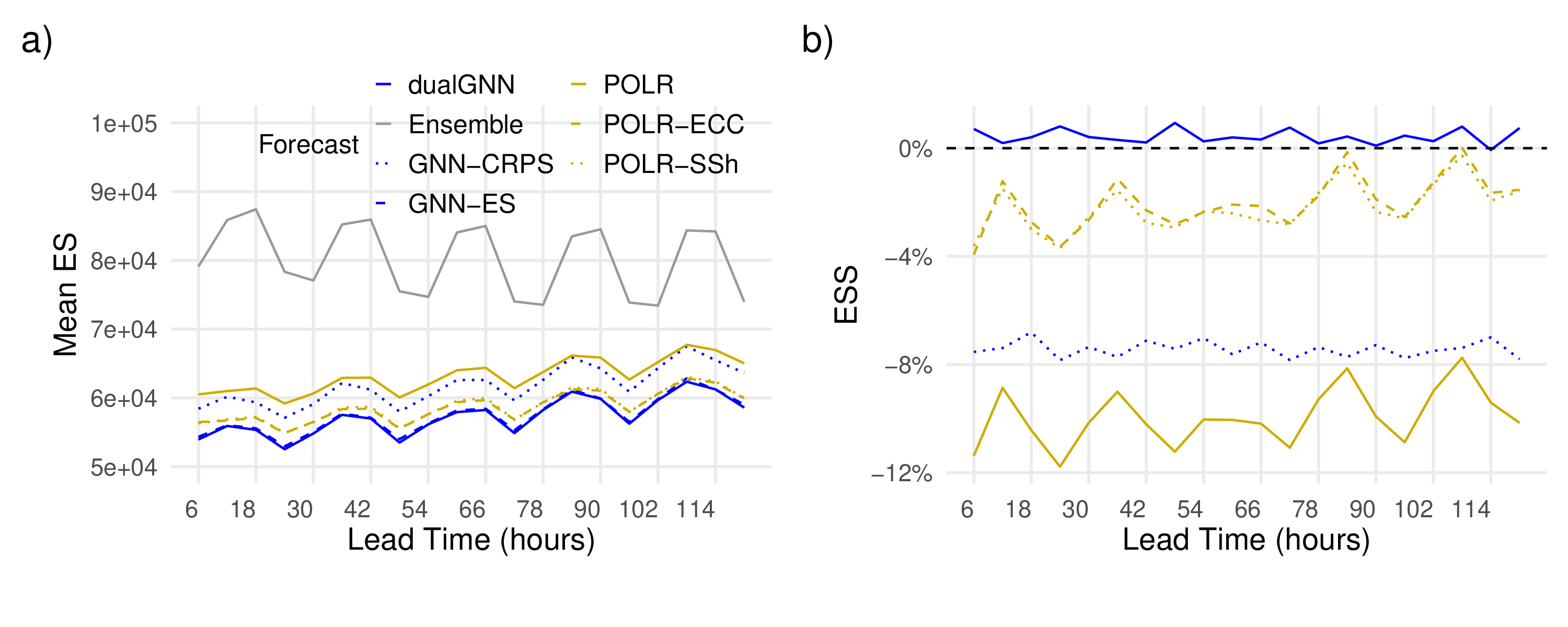}
    \caption{(a) Mean ES of raw, POLR, and multivariate post-processed visibility forecasts; (b) ESS of post-processed forecasts relative to GNN-ES. Both shown as functions of the lead time.}
    \label{fig:es_ess_vis}
\end{figure}

In Figure \ref{fig:es_dm}b, a somewhat different picture emerges. In this case, dualGNN outperforms the ES-trained GNN in 46.15\% of cases, while in 50\% of cases the null hypothesis of equal predictive performance cannot be rejected. For all MLP variants, including MLP-GNN, the improvement is limited to only 3.84\%; among these, MLP-GNN most frequently achieves performance that is statistically comparable to the baseline. GNN-CRPS follows in the ranking: although it never surpasses the ES-trained model, its performance coincides with it in 79.92\% of cases. By contrast, all EMOS variants are consistently outperformed by the ES-trained GNN for all lead times.

As shown in Figure \ref{fig:vs_dm}a, which evaluates the significance of the VS deviations relative to the strongest GNN-independent MLP-SSh reference forecast, the dualGNN again yields the largest gains over the strongest two-step benchmark, surpassing it in 76.92\% of cases. The MLP-GNN ranks second, showing improvements in 65.38\% of cases; notably, the benchmark model is never outperformed by either model.  These forecasts are followed by GNN-ES (46.15\%), MLP (7.69\%), and MLP-ECC (3.85\%). By contrast, GNN-CRPS and all EMOS variants never achieve significantly better performance than MLP-SSh; moreover, GNN-CRPS attains statistically indistinguishable skill in 92.31\% of cases, whereas the EMOS models, in line with earlier findings, are consistently outperformed.

\begin{figure}[t!]
    \centering    \includegraphics[width=1\linewidth]{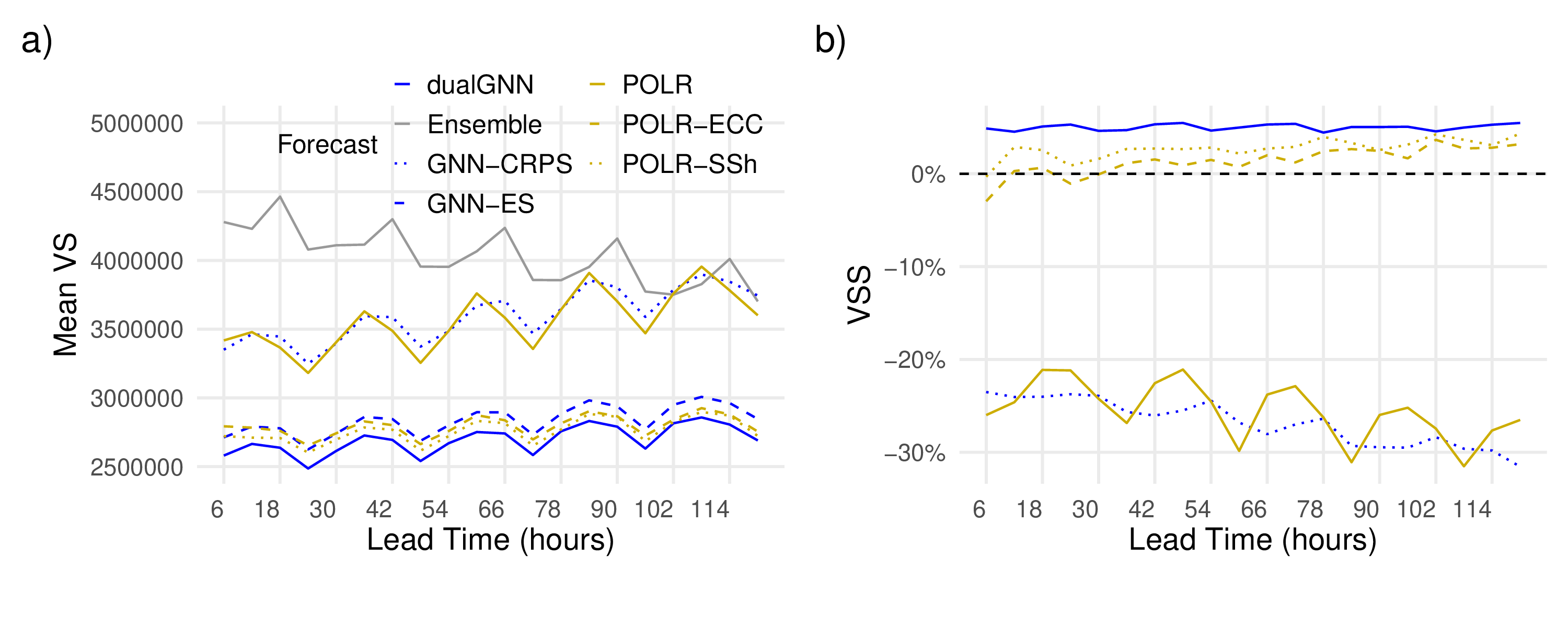}
    \caption{(a) Mean VS of raw, POLR, and multivariate post-processed visibility forecasts; (b) VSS of post-processed forecasts relative to GNN-ES.
Both shown as functions of the lead time.}
    \label{fig:vs_vss_vis}
\end{figure}

In Figure \ref{fig:vs_dm}b, where GNN-ES serves as the baseline, dualGNN achieves improvements in 84.61\% of cases and never performs worse, indicating stable improvements across lead times. The MLP variants follow, with MLP-GNN yielding the highest improvement rate (15.38\%), and the others achieving 11.54\%. Similar to Figure \ref{fig:vs_dm}a, EMOS variants and GNN-CRPS never surpass the baseline. In addition, the proportion of indistinguishable performance for GNN-CRPS decreases to 61.54\%.

Figure \ref{fig:mv_hists_rad} presents the multivariate rank histograms of the post-processed forecasts, with the associated RIs indicated on the plots; the lowest RIs are highlighted in bold. While the RI provides a convenient summary of deviations from uniformity, interpreting multivariate rank histograms requires caution, as their meaning depends on the choice of pre-ranking method. In general, the two-step post-processing models and the GNN-based approaches produce the most uniform average- and band-depth rank histograms, with dualGNN exhibiting the most even distributions and the independent EMOS model the highest RI. All models show a left-skew in the Energy Score rank histograms, with the effect being least pronounced for GNN-ES. Based on the VS-based Dependence Rank histograms, the histogram corresponding to the MLP reordered according to the GNN rank structure is the most uniform, whereas deviations in the correlation structure would manifest as right- or left-skewed histograms.

\subsection{Multivariate performance of visibility forecasts}
\label{sec:mv_results_vis}

The following section focuses on the multivariate post-processing of visibility forecasts. In addition to the POLR and GNN samples introduced in Section \ref{sec:uv_results_vis}, we consider the POLR-ECC variant, obtained by reordering the POLR samples according to the rank structure of the raw ECMWF ensemble, and the POLR-SSh models, rearranged based on the rank structure of past observations, as described in Section \ref{sec:uv_results_vis}. Similarly, to the previously analyzed variable, the SSh models employ observations from the training period, to determine the dependence template.

\begin{figure}
    \centering    
    \includegraphics[width=1\linewidth]{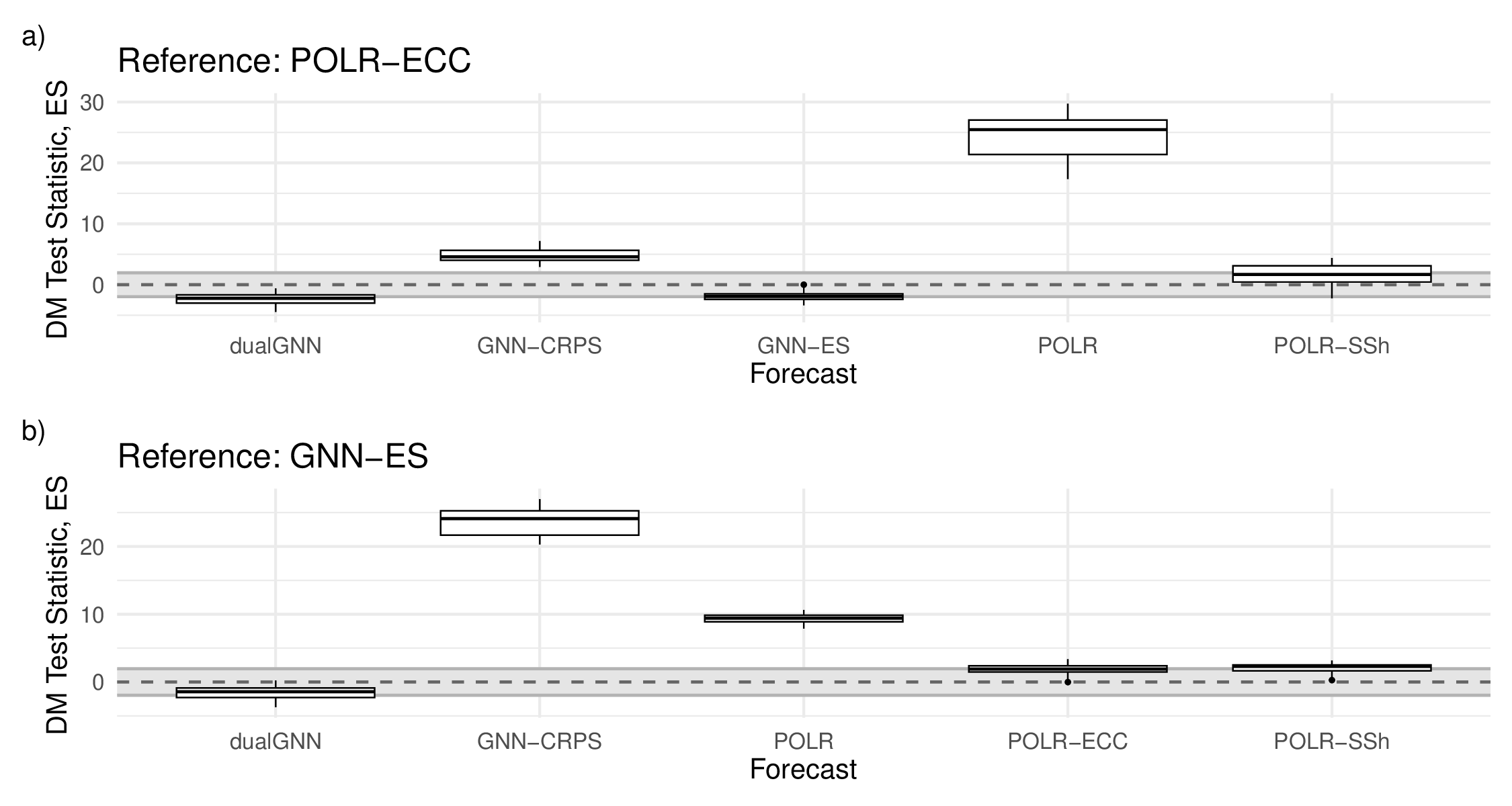}
    \caption{Boxplots of Diebold–Mariano (DM) test statistics investigating the significance of the difference in ES  (a) from the reference POLR-ECC and (b) GNN-ES methods as functions of the lead time. Grey region indicates the acceptance region of the two-tailed DM test for equal predictive performance at a 5\% level of significance.}
    \label{fig:es_dm_vis}
\end{figure}

Figure \ref{fig:es_ess_vis} shows the mean ES and the corresponding skill scores of POLR, and multivariate post-processed visibility forecasts relative to GNN-ES, with the raw forecasts omitted from the skill score plot to better highlight the differences between the post-processed models. As illustrated in Figure \ref{fig:es_ess_vis}a, all post-processed forecasts improve upon the raw ensemble predictions, although this advantage diminishes with increasing lead time. In order to highlight differences between post-processing models, in Figure \ref{fig:es_ess_vis}b the GNN-ES is used as the reference. Relative to this benchmark, dualGNN shows only a 0.4\% improvement, and it is the only model with a positive skill score; nevertheless, the performance of dualGNN and GNN-ES can be considered effectively equivalent. The strongest POLR variant, POLR-ECC, lags slightly behind the reference by 2\%, closely followed by POLR-SSh, representing a further 0.2\% decrease in the mean skill score. The weakest post-processing model is POLR, which exhibits an average 9.94\% decrease relative to GNN-ES.

A slightly different pattern observed in Figure \ref{fig:vs_vss_vis}, which displays the mean VS of POLR, and multivariate post-processed visibility forecasts, along with the corresponding skill scores, again using GNN-ES as the reference, with the raw forecasts omitted from Figure \ref{fig:vs_vss_vis}b. Unlike in Figure \ref{fig:es_ess_vis}a, clear improvements over the raw ensemble are now observed only for the GNN models that minimize multivariate scores and the two-step POLR methods. As before, this advantage diminishes with increasing lead time. Based on the mean VS, the models are ranked as follows: dualGNN, POLR-SSh, POLR-ECC, GNN-ES, POLR, GNN-CRPS, and finally the raw ensemble.

Figure \ref{fig:vs_vss_vis}b allows quantifying the improvement of the models relative to the ES-minimizing GNN. On average, the GNN trained to minimize CRPS performs 26.83\% worse, while the baseline POLR model exhibits a similar decrease of 25.51\%. Among the post-processing approaches, POLR-ECC is the first to achieve a positive gain relative to GNN-ES, with an improvement of 1.37\%, followed by POLR-SSh with 2.73\%. The largest gain is observed for dualGNN, which outperforms GNN-ES by 5\%.

\begin{figure}[t!]
    \centering    
    \includegraphics[width=1\linewidth]{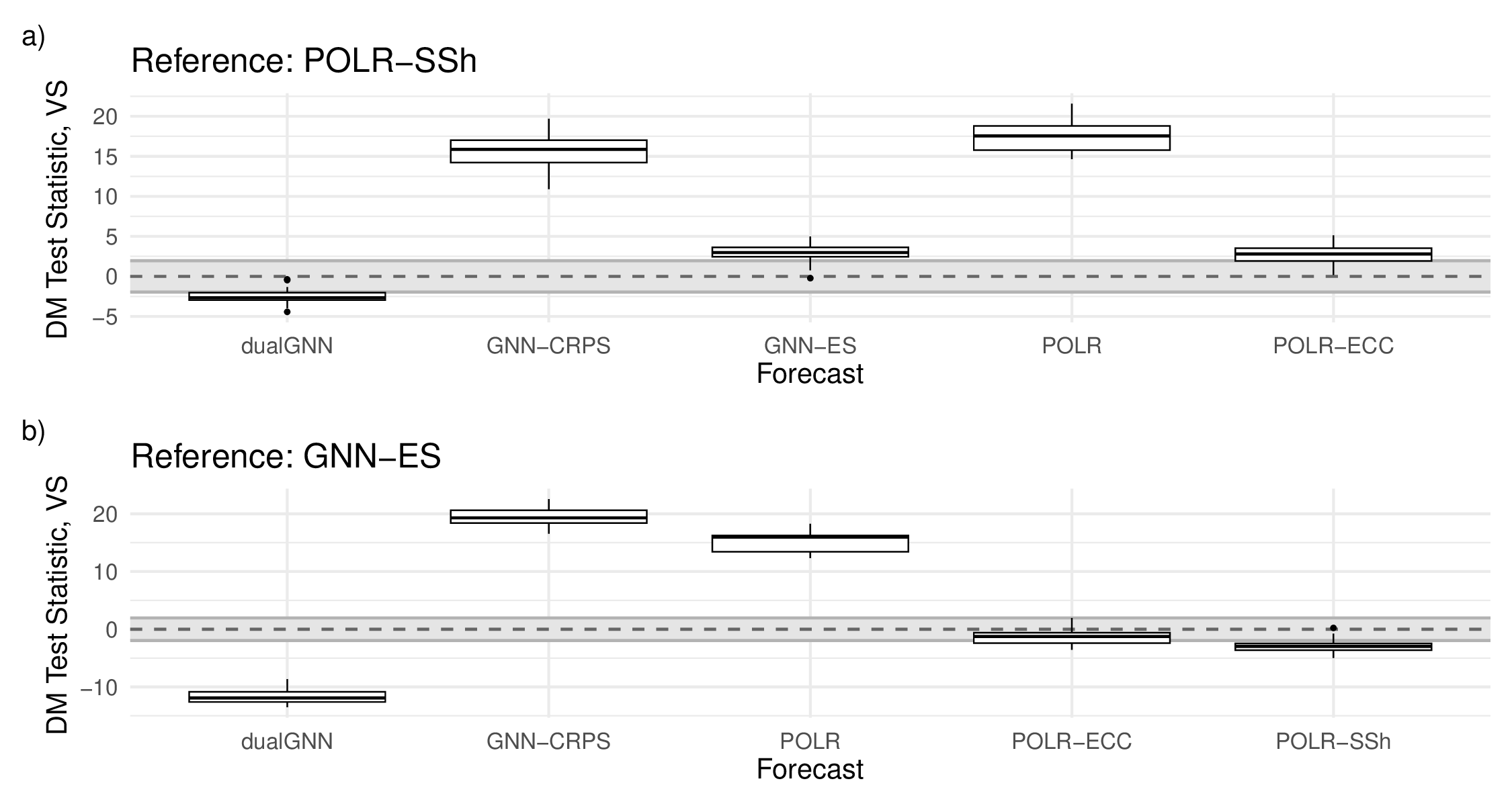}
    \caption{Boxplots of Diebold–Mariano (DM) test statistics investigating the significance of the difference in VS  (a) from the reference POLR-ECC and (b) GNN-ES methods as functions of the lead time. Grey region indicates the acceptance region of the two-tailed DM test for equal predictive performance at a 5\% level of significance.}
    \label{fig:vs_dm_vis}
\end{figure}

Figure \ref{fig:es_dm_vis} presents boxplots of DM test statistics, assessing the significance of differences in ES relative to POLR-ECC, the strongest GNN-independent two-step model, and GNN-ES, across different lead times. The grey area marks the acceptance region of the two-tailed DM test at the 5\% significance level, and, as in previous analyses, the Benjamini–Hochberg procedure was applied to control the false discovery rate. 

As shown in Figure \ref{fig:es_dm_vis}a, only GNN-ES and dualGNN significantly outperform the reference POLR-ECC, with improvements observed in 10\% and 45\% of lead times, respectively. For the remaining lead times, the null hypothesis of equal performance cannot be rejected, making these the only models that surpass POLR-ECC. Consistent with the findings from Figure \ref{fig:es_ess_vis}, GNN-CRPS and POLR remain below the reference, while POLR-SSh achieves comparable performance in 55\% of lead times but falls short for the rest.

\begin{figure}[t!]
    \centering    
    \includegraphics[width=1\linewidth]{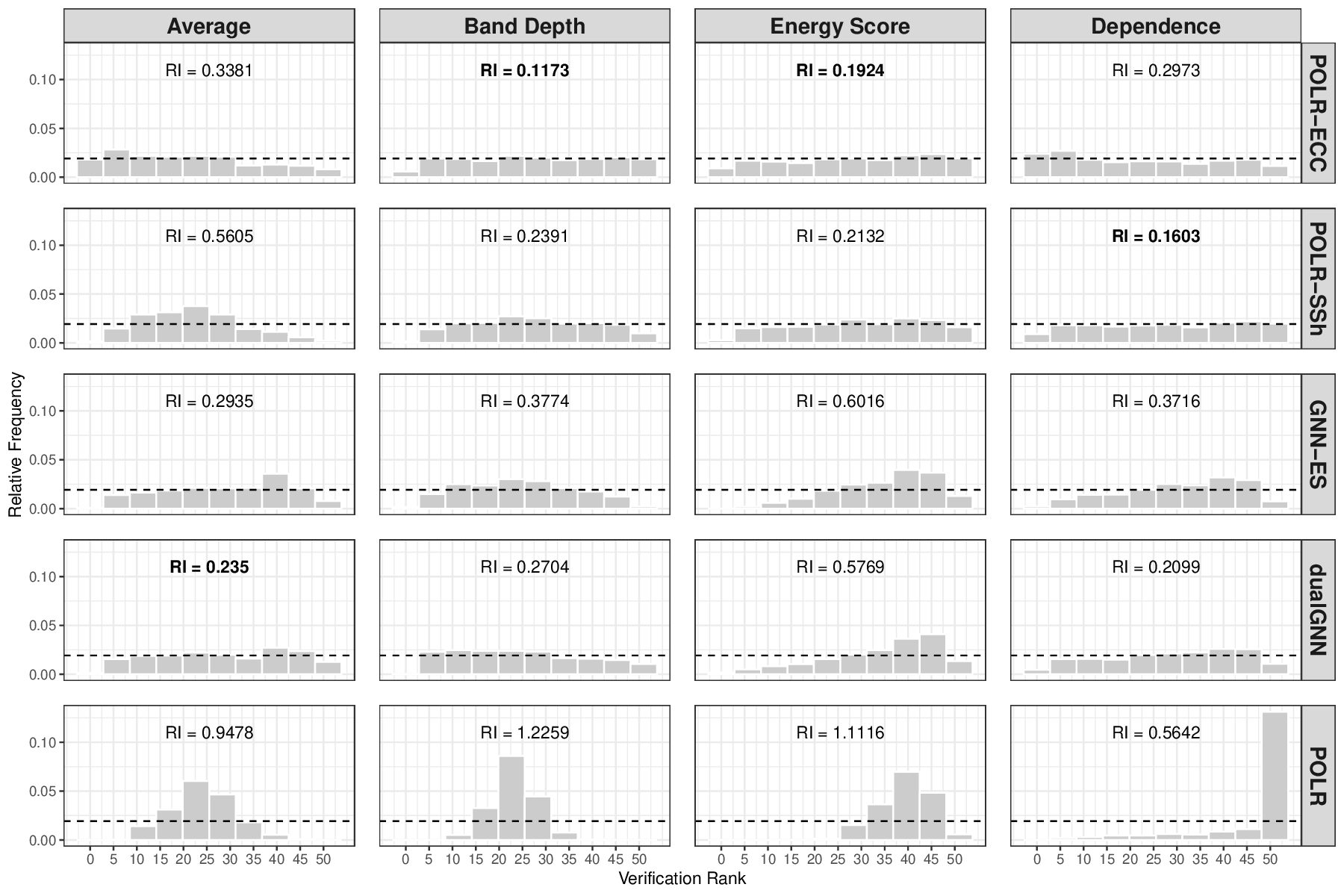}
    \caption{Rank histograms of post-processed visibilty forecasts together with the corresponding reliability indices}
    \label{fig:mv_hists_vis}
\end{figure}

In Figure \ref{fig:es_dm_vis}b, we again adopt the ES-minimizing GNN as the reference model. Relative to this benchmark, only dualGNN shows an improvement, outperforming GNN-ES in 25\% of lead times and matching its performance in the remaining cases, making it the only model with consistent gains. POLR-ECC follows, equaling GNN-ES in 90\% of lead times and underperforming in the remainder, with POLR-SSh trailing behind. As in previous analyses, GNN-CRPS and POLR never surpass the reference.

Figure \ref{fig:vs_dm_vis} presents boxplots of DM test statistics, evaluating the significance of differences in VS relative to POLR-SSh, the strongest GNN-independent two-step model, and GNN-ES, across various lead times. The grey area indicates the acceptance region of the two-tailed DM test at the 5\% significance level, and, as in previous analyses, adjustments for multiple comparisons were made using the Benjamini–Hochberg procedure to control the false discovery rate.

As shown in Figure \ref{fig:vs_dm_vis}a, no model outperforms the reference based on the past-observations-derived dependence template, except for dualGNN, which performs better in 70\% of lead times and matches the reference in the remaining 30\%. Among the other models, only POLR-ECC and GNN-ES come close to the performance of POLR-SSh, with GNN-ES equaling POLR-SSh in 20\% of lead times and POLR-ECC in 25\%, while underperforming in the remainder.

Figure \ref{fig:vs_dm_vis}b demonstrates the consistently strong performance of dualGNN, where relative to the GNN-ES, dualGNN exhibits a statistically significant improvement in all lead times, followed by POLR-SSh in 80\% of lead times and POLR-ECC in 25\% of lead times. In contrast, POLR and GNN-CRPS persistently underperform relative to the reference.

Finally, the improved calibration of the post-processed models can be further examined in Figure \ref{fig:mv_hists_vis}, although the histograms appear somewhat more complex compared to Figure \ref{fig:mv_hists_rad}. Based on the average ranks, dualGNN seems to perform the best, while the Energy Score and Dependence histograms suggest a tendency toward overestimated correlation. It is also notable that, in the case of the histograms, the GNN models are not constrained by the maximum generatable value as the POLR model is, which may make them more sensitive to certain patterns.

\section{Discussion}

The aim of this study was the multivariate post-processing of ensemble forecasts for different weather quantities, specifically solar irradiance and visibility, using a graph neural network trained with a combination of the Energy Score (ES) and the Variogram Score (VS). The primary objective was to demonstrate that an optimal combination of ES and VS can outperform a GNN trained solely on ES, as well as traditional two-step methods based on empirical copulas.

For both weather variables, the composite-loss GNN (dualGNN) consistently improved upon all reference models, achieving the best performance according to multivariate scores while maintaining well-calibrated marginal distributions. In contrast, empirical copula-based methods did not yield a single clear winner, whereas dualGNN achieved the best performance across both ES and VS. In the case of visibility forecasts, dualGNN also achieved the lowest mean CRPS, surpassed only by a GNN specifically optimized for CRPS. Furthermore, the rank structure of forecasts generated by dualGNN captured relevant dependencies among stations, producing the most uniform rank histograms. Being nonparametric, the method is applicable to any weather variable and allows flexibility in the number of ensemble members generated. Overall, the dualGNN consistently generated accurate forecasts across target variables with diverse characteristics. This demonstrates that the model can robustly handle variables with markedly different statistical properties while sustaining high predictive skill and preserving meaningful rank dependencies.

Despite these advantages, the dualGNN has limitations, as is often the case for many machine-learning methods compared to classical statistical models. One such limitation is increased data requirements, which were particularly evident for visibility forecasts. While an optimal training period could be identified for solar irradiance with the available data, visibility post-processing would likely benefit from additional training data. However, to maintain a reasonably long verification period, the training period selected here was used in this study.

As with any spatial GNN application, the construction of the graph is crucial. A 50 km threshold for defining edges between stations in northern Chile represents a reasonable choice, although finer graph structures may be more appropriate for visibility forecasts. In the current datasets, clustering stations based on climatological or geographical similarity followed by subsequent edge definition did not provide additional benefits. Nevertheless, a principal advantage of the multivariate loss function lies in the ability of the GNN architecture to produce calibrated samples, even when the underlying graphs are imperfectly specified, while simultaneously capturing dependency structures through joint optimization.

Future research could explore the inclusion of additional variables correlated with the target variable as features, potentially enabling richer feature sets and more informative graph structures, thereby improving performance. Alternative graph architectures, such as graph attention networks or graph convolutional networks, could also be investigated. 

An additional benefit of the GraphSAGE architecture used here is its scalability for large graphs, making it promising for future work on a global network of stations, as well as for comparison with other multivariate score-based models \citep{chen2024}. Moreover, one of the most promising aspects of dualGNN is its potential for spatial interpolation. However, for variables measured across highly heterogeneous environments, such as the diverse elevations and coastal locations in Chile, a substantially richer graph structure would be required. Preliminary tests suggest that interpolation can yield up to a 20\% improvement for certain stations without observations, while locations with significantly different climatological characteristics require further investigation. Comparative studies with spatial interpolation models such as of \cite{baran2024clustering} would also be of interest.

\bigskip
\noindent
{\bf Acknowledgments.} \ %This work was supported by the EK{\"O}P-25-4-II University Research Scholarship Program of the Ministry for Culture and Innovation, funded by the National Research, Development and Innovation Fund.  
The author gratefully acknowledges the support of the  National Research, Development, and Innovation Office under Grant No. K142849. Furthermore, she thanks her former supervisor, Sándor Baran, for kindly reviewing and providing valuable feedback on the manuscript.

\newpage

\bibliographystyle{apalike}

\begin{thebibliography}{}

\bibitem[Allen et~al., 2024]{allen2024assessing}
Allen, S., Ziegel, J., and Ginsbourger, D. (2024).
\newblock Assessing the calibration of multivariate probabilistic forecasts.
\newblock {\em Q. J. R. Meteorol. Soc.}, 150(760):1315--1335.

\bibitem[Baran and Baran, 2024]{bb24}
Baran, {\'A}. and Baran, S. (2024).
\newblock A two-step machine-learning approach to statistical post-processing
  of weather forecasts for power generation.
\newblock {\em Q. J. R. Meteorol. Soc.}, 150(759):1029--1047.

\bibitem[Baran and Lakatos, 2024]{baran2024clustering}
Baran, S. and Lakatos, M. (2024).
\newblock Clustering-based spatial interpolation of parametric postprocessing
  models.
\newblock {\em Wea. Forecasting}, 39(11):1591--1604.

\bibitem[Baran et~al., 2025]{bmcdsznl25}
Baran, S., Mar\'{\i}n, J.~C., Cuevas, O., D\'{\i}az, M., Szab\'o, M., Nicolis,
  O., and Lakatos, M. (2025).
\newblock Machine-learning-based probabilistic forecasting of solar irradiance
  in \text{Chile}.
\newblock {\em Adv. Stat. Climatol. Meteorol.}, 11(1):89--105.

\bibitem[Benjamini and Hochberg, 1995]{benjamini1995controlling}
Benjamini, Y. and Hochberg, Y. (1995).
\newblock Controlling the false discovery rate: a practical and powerful
  approach to multiple testing.
\newblock {\em J. R. Stat. Soc. Ser. B (Methodol.)}, 57(1):289--300.

\bibitem[Bouall{\`e}gue et~al., 2024]{bouallegue2024improving}
Bouall{\`e}gue, Z.~B., Weyn, J.~A., Clare, M.~C., Dramsch, J., Dueben, P., and
  Chantry, M. (2024).
\newblock Improving medium-range ensemble weather forecasts with hierarchical
  ensemble transformers.
\newblock {\em Artif. Intell. Earth Syst.}, 3(1):e230027.

\bibitem[B{\"u}lte et~al., 2025]{bulte2025graph}
B{\"u}lte, C., Maskey, S., Scholl, P., von Berg, J., and Kutyniok, G. (2025).
\newblock Graph neural networks for enhancing ensemble forecasts of extreme
  rainfall.
\newblock {\em arXiv preprint arXiv:2504.05471}.

\bibitem[Chen et~al., 2024]{chen2024}
Chen, J., Janke, T., Steinke, F., and Lerch, S. (2024).
\newblock Generative machine learning methods for multivariate ensemble
  postprocessing.
\newblock {\em Ann. Appl. Stat.}, 18(1):159--183.

\bibitem[Clark et~al., 2004]{clark2004schaake}
Clark, M., Gangopadhyay, S., Hay, L., Rajagopalan, B., and Wilby, R. (2004).
\newblock The \text{Schaake} shuffle: A method for reconstructing space--time
  variability in forecasted precipitation and temperature fields.
\newblock {\em J. Hydrometeorol.}, 5(1):243--262.

\bibitem[Dai and Hemri, 2021]{dai2021spatially}
Dai, Y. and Hemri, S. (2021).
\newblock Spatially coherent postprocessing of cloud cover ensemble forecasts.
\newblock {\em Mon. Weather Rev.}, 149(12):3923--3937.

\bibitem[Diebold and Mariano, 1995]{diebold2002comparing}
Diebold, F.~X. and Mariano, R.~S. (1995).
\newblock Comparing predictive accuracy.
\newblock {\em J. Bus. Econ. Stat.}, 13:253–263.

\bibitem[Feik et~al., 2024]{feik2024graph}
Feik, M., Lerch, S., and St{\"u}hmer, J. (2024).
\newblock Graph neural networks and spatial information learning for
  post-processing ensemble weather forecasts.
\newblock {\em arXiv preprint arXiv:2407.11050}.

\bibitem[Fey and Lenssen, 2019]{fey2019fast}
Fey, M. and Lenssen, J.~E. (2019).
\newblock Fast graph representation learning with \text{PyTorch Geometric}.
\newblock {\em arXiv preprint arXiv:1903.02428}.

\bibitem[Gneiting and Raftery, 2007]{gneiting2007strictly}
Gneiting, T. and Raftery, A.~E. (2007).
\newblock Strictly proper scoring rules, prediction, and estimation.
\newblock {\em J. Amer. Statist. Assoc.}, 102:359--378.

\bibitem[Gneiting et~al., 2005]{g2005}
Gneiting, T., Raftery, A.~E., Westveld~III, A.~H., and Goldman, T. (2005).
\newblock Calibrated probabilistic forecasting using ensemble model output
  statistics and minimum crps estimation.
\newblock {\em Mon. Weather Rev.}, 133(5):1098--1118.

\bibitem[Hamill and Colucci, 1997]{hamill1997verification}
Hamill, T.~M. and Colucci, S.~J. (1997).
\newblock Verification of \text{Eta}--\text{RSM} short-range ensemble
  forecasts.
\newblock {\em Mon. Weather Rev.}, 125(6):1312--1327.

\bibitem[Hamilton et~al., 2017]{hamilton2017inductive}
Hamilton, W., Ying, Z., and Leskovec, J. (2017).
\newblock Inductive representation learning on large graphs.
\newblock {\em Advances in neural information processing systems}, 30.

\bibitem[Jiang and Luo, 2022]{JIANG2022117921}
Jiang, W. and Luo, J. (2022).
\newblock \text{Graph neural network for traffic forecasting: A survey}.
\newblock {\em Expert Syst. Appl.}, 207:117921.

\bibitem[Knüppel et~al., 2022]{knuppel2022}
Knüppel, M., Krüger, F., and Pohle, M. O. (2022).
\newblock Score-based calibration testing for multivariate forecast distributions.
\newblock {\em arXiv preprint arXiv:2211.16362}.

\bibitem[Lakatos and Baran, 2024]{lakatos2024enhancing}
Lakatos, M. and Baran, S. (2024).
\newblock Enhancing multivariate post-processed visibility predictions
  utilizing copernicus atmosphere monitoring service forecasts.
\newblock {\em Meteorol. Appl.}, 31(6):e70015.

\bibitem[Lakatos et~al., 2023]{lakatos2023comparison}
Lakatos, M., Lerch, S., Hemri, S., and Baran, S. (2023).
\newblock Comparison of multivariate post-processing methods using global
  \text{ECMWF} ensemble forecasts.
\newblock {\em Q. J. R. Meteorol. Soc.}, 149(752):856--877.

\bibitem[Lerch and Baran, 2017]{lerch2017similarity}
Lerch, S. and Baran, S. (2017).
\newblock Similarity-based semilocal estimation of post-processing models.
\newblock {\em J. R. Stat. Soc. Ser. C Appl. Stat.}, 66(1):29--51.

\bibitem[Li et~al., 2025]{li2025kolmogorov}
Li, L., Zhang, Y., Wang, G., and Xia, K. (2025).
\newblock Kolmogorov--\text{Arnold} graph neural networks for molecular
  property prediction.
\newblock {\em Nat. Mach. Intell.}, 7:1346--1354.

\bibitem[Li et~al., 2022]{li2022convolutional}
Li, W., Pan, B., Xia, J., and Duan, Q. (2022).
\newblock Convolutional neural network-based statistical post-processing of
  ensemble precipitation forecasts.
\newblock {\em J. Hydrol.}, 605:127301.

\bibitem[McCullagh, 1980]{mccullagh1980regression}
McCullagh, P. (1980).
\newblock Regression models for ordinal data (with discussion).
\newblock {\em J. R. Stat. Soc. Series B Stat. Methodol.}, 42:243--268.

\bibitem[M{\"o}ller et~al., 2013]{moller2013multivariate}
M{\"o}ller, A., Lenkoski, A., and Thorarinsdottir, T.~L. (2013).
\newblock Multivariate probabilistic forecasting using ensemble bayesian model
  averaging and copulas.
\newblock {\em Q. J. R. Meteorol. Soc.}, 139(673):982--991.

\bibitem[Pic et~al., 2025]{pic2025distributional}
Pic, R., Dombry, C., Naveau, P., and Taillardat, M. (2025).
\newblock Distributional regression U-Nets for the postprocessing of
  precipitation ensemble forecasts.
\newblock {\em Artif. Intell. Earth Syst.}, 4(4):240067

\bibitem[Raftery et~al., 2005]{r200}
Raftery, A.~E., Gneiting, T., Balabdaoui, F., and Polakowski, M. (2005).
\newblock Using bayesian model averaging to calibrate forecast ensembles.
\newblock {\em Mon. Weather Rev.}, 133(5):1155--1174.

\bibitem[Rasp and Lerch, 2018]{rl18}
Rasp, S. and Lerch, S. (2018).
\newblock Neural networks for postprocessing ensemble weather forecasts.
\newblock {\em Mon. Weather Rev.}, 146(11):3885--3900.

\bibitem[Schefzik et~al., 2013]{schefzik2013uncertainty}
Schefzik, R., Thorarinsdottir, T.~L., and Gneiting, T. (2013).
\newblock Uncertainty quantification in complex simulation models using
  ensemble copula coupling.
\newblock {\em Stat. Sci.}, 28(4):616–640.

\bibitem[Scheuerer and Hamill, 2015]{scheuerer2015variogram}
Scheuerer, M. and Hamill, T.~M. (2015).
\newblock Variogram-based proper scoring rules for probabilistic forecasts of
  multivariate quantities.
\newblock {\em Mon. Weather Rev.}, 143(4):1321--1334.

\bibitem[Schulz et~al., 2021]{schulz2021}
Schulz, B., El~Ayari, M., Lerch, S., and Baran, S. (2021).
\newblock Post-processing numerical weather prediction ensembles for
  probabilistic solar irradiance forecasting.
\newblock {\em Sol. Energy}, 220:1016--1031.

\bibitem[Skamarock et~al., 2019]{skamarock2019}
Skamarock, W.~C., Klemp, J.~B., Dudhia, J., Gill, D.~O., Liu, Z., Berner, J.,
  Wang, W., Powers, J.~G., Duda, M.~G., Barker, D.~M., et~al. (2019).
\newblock A description of the advanced research wrf version 4.
\newblock {\em NCAR \text{Tech. Note NCAR/TN-556+STR}}, Available at:
  \url{https://api.semanticscholar.org/CorpusID:196211930} [Accessed on 31 July
  2025].

\bibitem[Thorarinsdottir and Gneiting, 2010]{thorarinsdottir2010probabilistic}
Thorarinsdottir, T.~L. and Gneiting, T. (2010).
\newblock Probabilistic forecasts of wind speed: ensemble model output
  statistics by using heteroscedastic censored regression.
\newblock {\em J. R. Stat. Soc. Ser. A Stat. Soc.}, 173(2):371--388.

\bibitem[Thorarinsdottir et~al., 2016]{thorarinsdottir2016assessing}
Thorarinsdottir, T.~L., Scheuerer, M., and Heinz, C. (2016).
\newblock Assessing the calibration of high-dimensional ensemble forecasts using rank histograms.
\newblock {\em J. Comput. Graph. Stat.}, 25(1):105--122.


\bibitem[Vannitsem et~al., 2021]{vannitsem2021pp}
Vannitsem, S., Bremnes, J.~B., Demaeyer, J., Evans, G.~R., Flowerdew, J.,
  Hemri, S., Lerch, S., Roberts, N., Theis, S., Atencia, A., et~al. (2021).
\newblock Statistical postprocessing for weather forecasts: Review, challenges,
  and avenues in a big data world.
\newblock {\em Bull. Am. Meteorol. Soc.}, 102(3):E681--E699.

\bibitem[Wilks, 2019]{wilks2019statistical}
Wilks, D.~S. (2019).
\newblock {\em Statistical Methods in the Atmospheric Sciences}.
\newblock Elsevier, Amsterdam, 4th edition.

\bibitem[{World Meteorological Organization}, 2018]{wmo2018guide}
{World Meteorological Organization} (2018).
\newblock {\em Guide to Instruments and Methods of Observation. Volume I --
  Measurement of Meteorological Variables (WMO-No. 8)}.
\newblock WMO, Geneva.


\end{thebibliography}

\end{document}